\documentclass[12pt,epsf,amstex]{article}
\usepackage [dvips]{graphicx}
\usepackage{amsmath}
\usepackage{amssymb}
\usepackage{upgreek}
\usepackage{epsfig}
\usepackage{verbatim}

\addtocounter{secnumdepth}{1}
\usepackage[utf8]{inputenc}  
\usepackage[T1]{fontenc}

\usepackage{pdfsync}

\begin{document}

\newcommand{\bTheta}{\Theta}
\newcommand{\bjm}{\bar{\jmath}}
\newcommand{\jinst}{j}
\newcommand{\J}{J}
\newcommand{\vecQ}{\vec{Q}}
\newcommand{\MTh}{{M_{\scriptscriptstyle{\text{\mdseries th}}}}}
\newcommand{\calMTh}{{{\cal M}_{\scriptscriptstyle{\text{\mdseries th}}}}}
\newcommand{\Zq}{\Pi_q}

\newcommand{\hP}{\widehat{P}}

\newcommand{\bnu}{\bar{\nu}}
\newcommand{\bk}{{\sf k}}

\newcommand{\vecp}{\vec{p}}

\newcommand{\tC}{\widetilde{C}}
\newcommand{\tD}{\widetilde{D}}
\newcommand{\C}{C}
\newcommand{\D}{D}
\newcommand{\Cstar}{C_*}
\newcommand{\tP}{\widetilde{P}}
\newcommand{\tbeta}{\widetilde{\beta}}
\newcommand{\tcalM}{\widetilde{\cal M}}
\newcommand{\betast}{{\beta_*}}
\newcommand{\mbbX}{\mathbb{X}}

\newcommand{\dd}{\mbox{d}}
\newcommand{\ee}{\mbox{e}}

\newcommand{\wt}{\widetilde}
\newcommand{\wh}{\widehat}

\newcommand{\etaq}{\eta_q}
\newcommand{\etamq}{\eta_{-q}}
\newcommand{\etadq}{\eta^\dag_q}
\newcommand{\etadmq}{\eta^\dag_{-q}}
\newcommand{\xiq}{\xi_q}
\newcommand{\ximq}{\xi_{-q}}
\newcommand{\xidq}{\xi^\dag_q}
\newcommand{\xidmq}{\xi^\dag_{-q}}

\newcommand{\etadqq}{\eta^\dag_q\eta_q}
\newcommand{\etadmqmq}{\eta^\dag_{-q}\eta_{-q}}
\newcommand{\etaqmq}{\eta_q\eta_{-q}}
\newcommand{\etadqdmq}{\eta^\dag_q\eta^\dag_{-q}}
\newcommand{\xidqq}{\xi^\dag_q\xi_q}
\newcommand{\xidmqmq}{\xi^\dag_{-q}\xi_{-q}}
\newcommand{\xiqmq}{\xi_q\xi_{-q}}
\newcommand{\xidqdmq}{\xi^\dag_q\xi^\dag_{-q}}

\newcommand{\Etaone}{\etadqq + \etadmqmq - 1}
\newcommand{\Etatwo}{\etadqdmq + \etaqmq}
\newcommand{\Etafour}{\etadqq\etadmqmq}
\newcommand{\Etaoneplus}{\etadqq + \etadmqmq}
\newcommand{\Xione}{\xidqq + \xidmqmq - 1}
\newcommand{\Xitwo}{\xidqdmq + \xiqmq}
\newcommand{\Xifour}{\xidqq\xidmqmq}
\newcommand{\Xioneplus}{\xidqq + \xidmqmq}

\newcommand{\la}{\langle}
\newcommand{\ra}{\rangle}
\newcommand{\rac}{\rangle_{\rm c}}
\newcommand{\rc}{)}
\newcommand{\lc}{(}

\newcommand{\beq}{\begin{equation}}
\newcommand{\eeq}{\end{equation}}
\newcommand{\bea}{\begin{eqnarray}}
\newcommand{\eea}{\end{eqnarray}}

\numberwithin{equation}{section}

\thispagestyle{empty}
\title{\Large {\bf Glauber's Ising chain\\ between two thermostats\\
\phantom{xxx} }}
 
\author{{F.~Cornu and H.J.~Hilhorst}\\[5mm]
{\small Laboratoire de Physique Th\'eorique, B\^atiment 210,}\\[-1mm] 
{\small CNRS and Universit\'e Paris-Sud XI,}\\[-1mm]
{\small Universit\'e Paris-Saclay, 91405 Orsay Cedex, France}\\}

\maketitle

\begin{small}
\begin{abstract}
\noindent 
We consider a one-dimensional Ising model each of whose
$N$ spins is in contact with two
thermostats of distinct temperatures $T_1$ and $T_2$.
Under Glauber dynamics the stationary state happens to coincide with the
equilibrium state at an effective intermediate temperature $T(T_1,T_2)$.
The system nevertheless carries a nontrivial energy current between the thermostats.
By means of the fermionization technique, 
for a chain initially in equilibrium at an arbitrary temperature $T_0$
we calculate the Fourier transform of 
the probability  $P({\cal Q};\tau)$ for
the  time-integrated energy current ${\cal Q}$ during a finite time interval $\tau$.
In the long time limit we determine the corresponding generating function for the cumulants per site and unit of time
$\la{\cal Q}^n\ra_{\rm c}/(N\tau)$  
and explicitly exhibit those with $n=1,2,3,4.$
We exhibit various phenomena in specific regimes:  kinetic mean-field effects when one thermostat flips any spin less often  than the other one, as well as dissipation towards a thermostat at zero
temperature. Moreover, when the system size $N$ goes to infinity while  the effective temperature $T$ vanishes, the cumulants of ${\cal Q}$ per unit of time  grow linearly with $N$ and are equal to those of a random walk process.  In two adequate scaling regimes involving $T$ and  $N$ we exhibit the  dependence  of the first correction upon the ratio of the spin-spin correlation length $\xi(T)$ and the size $N$. 
 \\

\noindent
{{\bf Keywords:} driven Ising model, time-integrated energy flux, 
large deviation function, exact solution}
\end{abstract}
\end{small}
\vspace{10mm}

\noindent LPT Orsay 16/xx

\tableofcontents

\clearpage

\section{Introduction} 
\label{secintroduction}

Since a few decades  statistics of the currents that characterize an out-of equilibrium state have been intensely studied both  experimentally and theoretically. Indeed the fluctuations of these currents in small systems are non-negligible with respect to their mean value, and they now can be investigated at nano scale thanks to very fast technological improvements \cite{BustamanteETAL2005,Ritort2006}. Meanwhile, the theory of stochastic thermodynamics has been developed and  the large fluctuations of time-integrated currents in out-of-equilibrium systems  have  been shown to obey generic \textit{fluctuation relations}. The latter  have been derived under various hypotheses about the microscopic dynamics: deterministic or stochastic with either discrete or continuous degrees of freedom
\footnote{For a comprehensive review  see the report by Seifert Ref.\cite{Seifert2012} and the references therein. In particular, for the case of stochastic Markovian dynamics with jumps between a finite number of configurations see Ref.\cite{HarrisSchutz2007}.}.
These fluctuation relations for time-integrated currents quantify how the second law of thermodynamics, valid for mean currents, is modified at the scale of fluctuations; they are linked in some way to the fluctuations  of the  time-integrated entropy production rate in the system
\footnote{Short introductions which  point out the role of entropy  are  to be found \textit{e.g.} in Refs.\cite{Mallick2009,CornuBauer2013I,VanDenBroeck2013Varenna}.}. 
 In particular  the class of systems with a finite number of discrete degrees of freedom has provided firmly established fluctuation relations \cite{HarrisSchutz2007}.

Besides these generic fluctuation relations based on symmetry arguments,  solvable models have provided better insight into more detailed statistical properties of non-equilibrium stationary  states (NESS). This is most valuable 
in the absence of any equivalent of the equilibrium Gibbs ensemble theory for the description of NESS.
In particular two  paradigmatic kinetic models where a stationary  current  of particles or energy quanta  flows from one reservoir to another  have been widely investigated under various forms. On the one hand  one-dimensional systems of particles endowed with a simple exclusion process and non-equilibrium open boundary conditions; such models  describe particle exchange between two reservoirs  connected to both ends of the system and which have different chemical potentials (see reviews
\cite{ChouMallickZia2011,Mallick2015}). On the other hand Ising spin chains (with nearest-neighbor ferromagnetic interactions) where all spins are flipped by one of two thermostats.
\vskip 0.3 cm

In this paper we will introduce and study analytically a particuiar version of an Ising chain coupled to two thermostats. We begin by briefly recalling a few exact analytic results about kinetic Ising models.

In 1963 Glauber \cite{Glauber1963} endowed the Ising spin chain with a stochastic dynamics in order to describe the relaxation of this chain to its canonical equilibrium, which is  determined only by the Ising energy  and  a given temperature $T$.  A spin flip is interpreted as an energy exchange with a thermostat at temperature $T$. A  single spin is flipped  at a time, and the corresponding  Markovian  process is described by a master equation in spin configuration space. 

The relaxation to the canonical equilibrium is ensured by the choice of the transition rates made by Glauber: these are the simplest ones that obey the  detailed balance with the canonical configuration probability.  
The solution to the full description of the approach to equilibrium in this kinetic model was made in successive steps. First  Glauber  determined the  evolution of the average magnetization and spin-spin correlations, 
and studied the linear response to an applied
magnetic field.
In the early 1970's higher order correlation functions were studied
\cite{BedeauxEtAl1970,Felderhof1971-A}.
In particular, Felderhof \cite{Felderhof1971-A,Felderhof1971-B} was the first one to apply the fermionization technique to the Glauber model and showed
that the master equation is fully solvable: that is, for a system of
$N$ spins the $2^N$ eigenvalues and eigenvectors of the Markov matrix were all  found exactly.

Later kinetic models for  the Ising chain have been  introduced in order to investigate  the non-equilibrium stationary  state (NESS) sustained  by this  Ising chain when the spins are flipped by two thermostats at different temperatures. 

Exact results about the stationary  probability distribution of the spin configurations have been obtained through the determination of mean instantaneous quantities in various models \cite{RaczZia1994,MobiliaZiaSchmittmann2004,LavrentovichZia2010,Lavrentovich2012}. Analytical expressions for the large deviation function of the time-integrated energy current in the non-equilibrium stationary state (NESS) have been obtained for simpler  models \cite{FaragoPitard2007,FaragoPitard2008}. The complete  description of the time-integrated energy currents has been obtained for a model where   thermal contact between two thermostats is ensured by the interaction inside a set of independent Ising spin pairs, where each thermostat flips only one spin in the pair according to the corresponding Glauber dynamics \cite{CornuBauer2013}. The explicit  joint probability of the cumulative heats received from each thermostat at any time and the analytical expression for the large deviation function of the time-integrated heat transfer from one thermostat to the other were obtained 
\footnote{In the case of interacting Ising spin pairs one can obtain a partial description of  the energy transfer from one thermostat to the other: the generating function for the  long time cumulants per unit of time can be calculated analytically \cite{Cornu2017a}.
}. 
The explicit  stationary probability distributions of microscopic configurations have also been obtained  for other archetypal models: the asymmetric exclusion process \cite{Mallick2015} and several variants of the zero-range process \cite{EvansHanney2005,EvansWaclaw2014}. The generating function for the cumulants of the time-integrated particle current have  been obtained by sophisticated methods for various models endowed with an simple exclusion process\cite{Mallick2015}.
\vskip 0.3cm

In this work we study the Ising chain 
with a ferromagnetic nearest-neighbor coupling $E$,
a finite number $N$ of spins, and periodic
boundary conditions. The chain is
coupled to two thermostats at temperatures $T_1$ and $T_2$
in the simplest of all possible ways:
each spin may be reversed by either thermostat according to Glauber transition rates with inverse time constants (inverse time scales of random jumps) $\nu_1$ and $\nu_2$, respectively.
These are  kinetic parameters which depend on the microscopic dynamics of the system, as opposed to the thermodynamic parameters $T_1$ and $T_2$ of the energy reservoirs.
The amount of energy received by the chain for each spin flip is equal  to $-E$, $0$, or $+E$.
In the following all energies will be expressed as multiples of $4E$. We will take $T_1>T_2$ throughout this work. We rescale the physical time $t$ as $\tau=(\nu_1+\nu_2)t$ and the kinetic
parameters as $\bnu_a=\nu_a/(\nu_1+\nu_2)$, where $a=1,2$.

We are interested in the joint probability $P(Q_1,Q_2;\tau)$ for the stochastic energy amounts $Q_1$ and $Q_2$ received by the Ising chain from the thermostats during  a given time $\tau$. Then the probability for the time-integrated energy current ${\cal Q}$ (or net total energy that has flowed) from thermostat $1$ to thermostat $2$ during  time $\tau$ is obtained as the marginal probability for the variable ${\cal Q}=\tfrac{1}{2}(Q_1-Q_2)$. (We recall that $Q_a$ (with $a=1,2$) is an integer.)
\vskip0.3cm

The key to solvability is
the observation
\footnote{This observation goes back at least to Garrido {\it et al.}
\cite{GarridoEtAl1987}, whose focus is however different from ours.}
that the sum of two Glauber rates at temperatures $T_1$ and $T_2$ is a Glauber rate  with an effective kinetic parameter $\nu_1+\nu_2$ and at an intermediate temperature $T$ which is function of $T_1$, $T_2$ and $\bnu_1=\nu_1/(\nu_1+\nu_2)$. As a consequence, on the one hand, the transition rates obey the canonical  detailed balance and in a finite time the Ising spin chain reaches  its stationary state where the probability for a spin configuration is the Boltzmann-Gibbs weight at  the effective temperature $T(T_1,T_2,\bnu_1)$. Then the net instantaneous energy current on each site  has a zero mean, $\la \jinst \ra=0$, but the contribution to this mean current from each thermostat does not vanish, $\la \jinst_1 \ra= -\la \jinst_2 \ra \neq 0$.

In order to deal with the extended Markov matrix which governs the evolution of the Fourier transform of the joint probability $P(s,Q_1,Q_2;\tau)$ for spin configurations $s$ and exchanged quantities $Q_1$ and $Q_2$,
 we extend the 
 original method introduced by Felderhof \cite{Felderhof1971-A,Felderhof1971-B} for the Markov matrix of the probability $P(s;\tau)$ of the spin configurations during the  relaxation to equilibrium for an Ising chain coupled to a single thermostat.
This extended method yields 
all eigenvalues and eigenvectors of the extended master equation.
It allows us to calculate the Fourier transform of 
$P(Q_1,Q_2;\tau)$ and  $P({\cal Q};\tau)$
\footnote{We use the same symbol $P$ for various different probabilities; the meaning will always be clear.} 
at any time $\tau$.
The system fulfills the  hypotheses of  various generic  fluctuation relations,  (\ref{FRoneBis})-(\ref{FRoneS}) and (\ref{symmfNz})-(\ref{INfluctrel}), which are indeed satisfied by the  explicit  expressions for the involved quantities.

From the expression for the Fourier transform of the  probability  $P({\cal Q};\tau)$ of the time-integrated energy current ${\cal Q}=\tfrac{1}{2}(Q_1-Q_2)$, we obtain the explicit expression of the generating function for the infinite time limit of the  cumulants of ${\cal Q}$ per  site and unit of time, to be denoted as $\la{\cal Q}^n\rac/N\tau$. 
The $n$th cumulant (per site and unit of time) of interest, $\lim_{\tau\to\infty}\la{\cal Q}^n\rac/(N\tau)$, appears to be a $n$th degree 
polynomial in two variables ${\sf A}$ and ${\sf B}$
that are combinations of the
thermodynamic and kinetic parameters,
\beq
{\sf A} = \bnu_1\bnu_2(1-\gamma_1\gamma_2), \qquad
{\sf B} = \bnu_1\bnu_2(\gamma_2-\gamma_1),
\label{xABintro}
\eeq
where $\gamma_a = \tanh 2\beta_aE$ for $a=1,2$
\footnote{They are the same as the $A$ and $B$ of 
Ref.\,\cite{CornuBauer2013}, except that our
${\sf B}$ has a minus sign compared to $B$, 
due to an inversion of the roles of the two thermostats.}.
These polynomials have coefficients $\Sigma_n(N,\gamma)$ which depend on the system size $N$ and the  inverse effective temperature $\beta=(1/2E)\operatorname{artanh}\gamma$. They
 generalize the constant-coefficient polynomials
that appeared in work by Cornu and Bauer \cite{CornuBauer2013}
for a model where each thermostat flips only the spin on a given site. 
Although their model is different from the present chain with $N=2$,
its various symmetries render its energetics identical to that of the present $N=2$
system
\footnote{
Properties of their model  that are
invariant by a global spin flip are equivalent to the properties of our system
that are left-right invariant along the  chain with $N=2$.}.
\vskip0.3cm

The explicit solution for the long time cumulants per site and unit of time allows one to investigate several physical effects beyond the generic symmetry relations. Indeed kinetic and dissipation effects specific to various regimes of the thermodynamic and kinetic parameters can be investigated. They are summarized in the conclusion. 

Moreover  size effects generated by the interaction between spins can be  controlled. The model makes sense only if the effective temperature $\beta$ is finite ($\gamma\neq 1$). Then the  large deviation function exists in the infinite size limit and all long time cumulants per unit of time for the whole chain, $\lim_{\tau\to\infty}\la{\cal Q}^n\rac/\tau$,  are proportional to the size $N$ of the chain  at leading order in $N$. 
In  the double  limit where the effective temperature $1/\beta$ goes to zero while the size $N$ goes to infinity, all these cumulants are  proportional to $(1-\gamma)N$ at leading order in $N$ and $1-\gamma$. We notice that the factor $(1-\gamma)$ disappears if one considers the rescaled cumulants per unit of time when  the unit of time is the magnetization relaxation time $\tau_\textrm{rel}$, which is equal  to $[(\nu_1+\nu_2)(1-\gamma)]^{-1}$.
In this double limit  the  variables ${\sf A}$ and ${\sf B}$ defined in (\ref{xABintro}) vanish as $1-\gamma$ while the coefficients $\lim_{N\to\infty}\Sigma_n(N,\gamma)$ with $n\geq 2$ diverge.  
As a consequence, the leading behavior of the rescaled cumulants per unit of time is  a random walk contribution of order $N$, whereas the first correction to it  is not of order zero in $N$ when $1-\gamma\to 0$. In fact one has to consider two scaling regimes where the increase of $N\gg 1$ is related to the decrease of $1-\gamma\ll 1$;  we exhibit  how the first correction in the cumulants depends upon the ratio of the  spin-spin correlation length $\xi(T)$ and the size $N$.

This paper is set up as follows.
In section \ref{secdynIsing} we define the Ising model between two thermostats.
In section \ref{secdirect} we discuss  the instantaneous energy current,
whose average $\la j\ra$ per site we determine by elementary means.
In section \ref{secinjected} we define and diagonalize the master operator in
the extended space of spin configurations and energies $Q_1$ and $Q_2$ received by the spin chain from both thermostats during a time interval $\tau$, and
in section \ref{secaverages} we determine the Fourier transform of the joint probability  $P(Q_1,Q_2;\tau)$.  We check that the explicit expression of  $P(Q_1,Q_2;\tau)$ in the present model does satisfy  the fluctuation relations (\ref{FRoneBis})-(\ref{FRoneS}) which are retrieved from general considerations.
In section \ref{Qstat} we obtain the Fourier transform of the probability $P({\cal Q};\tau)$ of the time-integrated energy current 
${\cal Q}$ from one thermostat to the other during a time $\tau$.  We determine the cumulants per site and unit of time of ${\cal Q}$ in  the long-time limit  and discuss their structure. In section \ref{Physicaleffects}
 we study physical effects in  various regimes of the thermodynamic and kinetic parameters for a finite chain.  In  section \ref{secInfiniteSize} we consider a large size chain at very low effective temperature: from the study of some divergent coefficients performed  in Appendix \ref{Details} we exhibit the first correction to the leading $N$-behavior of the cumulants. In section \ref{secconclusion} we briefly conclude.

\section{Ising model coupled to two thermostats}
\label{secdynIsing}

We consider a chain of Ising spins $s_n=\pm1$, where $n=1,2,\ldots,N$ and 
$N\geq 2$ is an arbitrary integer. 
A configuration $s=(s_1,s_2,\ldots,s_{N})$ of the Ising model has 
an energy $H(s)$ given by
\beq
H(s)=-E\sum_{n=1}^{N} s_ns_{n+1}\,,
\label{defH}
\eeq
where we adopt the periodic boundary condition $s_{N+n}=s_n$.
We will be concerned with time dependent probability
 $P(s;\tau)$ in configuration space.

In a formalism that goes back at least to Kadanoff and Swift
\cite{KadanoffSwift1968} we associate
with each $s$ a ket $|s\ra =\otimes_{n=1}^N|s_n\ra$. 
A probability $P(s;\tau)$ is then
represented by a time dependent ket 
\beq
|P(\tau)\ra = \sum_{s}P(s;\tau)|s\ra.
\label{defketP}
\eeq
Since the classical discrete variables $s_n$ all commute, the Ising model has no dynamics of itself.
In 1963 Glauber \cite{Glauber1963} 
 stipulated that when the system is in contact with a thermostat at
temperature $T_1$, then in a configuration $s$ the spin $s_n$ on the $n$th 
lattice site may reverse its state with a transition  rate given in dimensionless time $\tau=(\nu_1+\nu_2) t$  (where $\nu_a$ is an inverse time) by 
\beq
w_n(s;\beta_1)=\frac{1}{2}\bnu_1[1-\tfrac{1}{2}\gamma_1 s_n(s_{n-1}+s_{n+1})],
\label{defwn}
\eeq
where
$\bnu_1=\nu_1/(\nu_1+\nu_2)$ is an inverse time,
$\gamma_1=\tanh 2\beta_1 E$, and $\beta_1=1/k_{\rm B}T_1$ is the inverse temperature.
The ket $|P(\tau)\ra$ then evolves according to the master equation
\beq
\frac{\dd}{\dd \tau} |P(\tau)\ra = \bnu_1\MTh(\beta_1)|P(\tau)\ra
\label{defmeq}
\eeq
with a ``master operator'' $\MTh(\beta_1)$ 
whose expression is originally due to Felderhof \cite{Felderhof1971-A,Felderhof1971-B},
\beq
\MTh(\beta_1) = \tfrac{1}{2}\sum_{n=1}^{N} (\sigma_n^x-1)
\big[1-\tfrac{1}{2}\gamma_1\,\sigma_n^z(\sigma_{n-1}^z+\sigma_{n+1}^z)\big],
\label{defMbeta}
\eeq
in which $\sigma_n^z$ and $\sigma_n^x$ are the usual
Pauli spin operators defined by
$\sigma_n^z|s_n\ra = s_n|s_n\ra$ and $\sigma_n^x|s_n\ra = |-s_n\ra$.
The master equation is easily shown to have the unique stationary state
\beq
|P_{\rm eq}(\beta_1)\ra = \rho_{\rm eq}(\beta_1)|1\ra,
\qquad |1\ra\equiv \sum_s|s\ra,
\label{statstate}
\eeq
in which we have
\beq
\rho_{\rm eq}(\beta_1)= \frac{\ee^{-\beta_1 {\cal H}}}{Z(\beta_1)}\,, \qquad 
{\cal H} =-E\sum_{n=1}^N \sigma_n^z\sigma_{n+1}^z\,, \qquad
Z(\beta_1)=\mbox{Tr}\,\ee^{-\beta_1 {\cal H}}.
\label{defcalP}
\eeq
We remark that $H(s)$ in Eq.\,(\ref{defH}) is an eigenvalue of ${\cal H}$.

By means of fermionization the operator $\MTh(\beta_1)$ may be completely
diagonalized and all its eigenvectors determined \cite{Felderhof1971-A,Felderhof1971-B}.
That means that, in principle, this problem is fully understood.
Recent renewal of interest in kinetic Ising models,
as mentioned in the introduction,
is due to the development of the study of non-equilibrium stationary state
systems.
With this perspective in mind
we will here couple the same system to two thermostats
at inverse temperatures $\beta_1$ and $\beta_2$
and acting with rates $\nu_1$ and $\nu_2$, respectively.
The total operator 
describing the system, denoted by $M$, then becomes a weighted sum of the Glauber
operators at inverse temperatures $\beta_1$ and $\beta_2$\,,
\beq
M=
\bnu_1\MTh(\beta_1) + \bnu_2\MTh(\beta_2).
\label{defMbis}
\eeq
with $\bnu_1+\bnu_2=1$.
In this work we study this model in detail.

Normally a system in contact with two reservoirs in different equilibrium states
will tend to a stationary state. Usually the precise properties
of such a state are not easy to determine. In the present case
a simplification occurs since
the operator $M$ of equation (\ref{defMbis}) can be rewritten as 
\beq
M =\MTh(\beta),
\label{exM}
\eeq
where $\beta$ represents an effective 
temperature intermediate between $\beta_1$ and $\beta_2$ given by
\beq
\tanh 2\beta E = \bnu_1\tanh 2\beta_1 E + \bnu_2\tanh 2\beta_2 E.
\label{defbeta}
\eeq
We will employ below the abbreviations 
$\gamma=\tanh 2\beta{E}$ 
and $\gamma_a=\tanh 2\beta_a{E}$ for $a=1,2$.

It follows that the stationary state in this case actually happens to be
equal to the equilibrium state at the effective temperature
\footnote{The same observation was made by Cornu and Bauer 
\cite{CornuBauer2013} for their two-spin system with only two energy levels.}.
This does not mean that we immediately 
know the answers to the questions raised above 
considering the energy injection and dissipation.
It means, however, that they can be calculated, which is what we do in this
work.

\section{Energy current between the thermostats}
\label{secdirect}

We consider the system in its stationary state, that is, in the equilibrium
state at inverse temperature $\beta$.
The reversal of a spin involves an energy change only if the two 
neighbors of that spin are mutually parallel. 
Let $f_{\rm al}$ be the fraction of all spins that have their
two neighbors mutually parallel and aligned to it, and $f_{\rm op}$ 
the fraction of those having them mutually parallel
and opposite to it. 
The indicator function for a spin $s_n$ aligned with (opposite to) both of its
neighbors is $\tfrac{1}{4}(1 \pm s_{n-1}s_n)(1 \pm s_ns_{n+1})$.
Ensemble averaging this by standard methods which lead to the result $\la s_n s_{n+r}\ra=[\zeta^r + \zeta^{N-r}]/[1+ \zeta^N]$, with $\zeta=\tanh\beta E$,  we obtain for the periodic
Ising chain
\beq
f_{\rm al,op} = 
\frac{1}{4}\left[ 1 \pm 2\frac{\zeta+\zeta^{N-1}}{1+\zeta^N} + 
\frac{\zeta^2+\zeta^{N-2}}{1+\zeta^N} \right],
\qquad N\geq 2.
\label{deffalop}
\eeq

We consider the action on this system by the operator $\bnu_1\MTh(\beta_1)$.
The spins of the two classes $f_{\rm al}$ and $f_{\rm op}$
are reversed with transition rates expressed in the dimensionless time $\tau=(\nu_1+\nu_2)t$ as
\beq
w_{\rm al,op}(\beta_1) = \tfrac{1}{2}\bnu_1(1 \mp \tanh 2\beta_1 E),
\label{defwalop}
\eeq
respectively. (The minus sign corresponds to $w_{\rm al}$.)
Let $\la \jinst_1\ra$ be the net average instantaneous energy current per unit of chain length
from thermostat $1$ into the system. Expressed in units of $4E$ 
it reads
\bea
\la \jinst_1\ra &=& f_{\rm al}\,w_{\rm al}(\beta_1) 
             - f_{\rm op}\,w_{\rm op}(\beta_1) \nonumber\\[2mm]
&& = \frac{\bnu_1}{2} 
\left[ \frac{\zeta+\zeta^{N-1}}{1+\zeta^N} 
 - \tfrac{1}{2}(1+\zeta^2)\frac{1+\zeta^{N-2}}{1+\zeta^N}
\,\tanh 2\beta_1 E \right].
\label{xj1av}
\eea
A similar expression holds for the net average current $\la \jinst_2\ra$ 
from thermostat $2$ into the system under the action of 
$\bnu_2\MTh(\beta_2)$. From (\ref{defbeta}) and (\ref{xj1av}) together with the relation 
$\tanh 2\beta E=\zeta^2/(1+\zeta^2)$, we get that 
$\la \jinst_1\ra  +\la \jinst_2\ra =0$  : in a stationary state 
the finite system cannot accumulate energy.
Then $ \la \jinst \ra =\la \jinst_1\ra = -\la \jinst_2\ra$
 represents the net average energy
current per site (= unit of chain length) that
traverses the system from thermostat $1$ to thermostat $2$.
The most elegant expression for this quantity is obtained by remembering that
$\bnu_1+\bnu_2=1$ and writing it as
$ \la \jinst \ra =\bnu_2\la \jinst_1\ra - \bnu_1\la \jinst_2\ra$ with the result
\beq
\la \jinst \ra = \tfrac{1}{4}\bnu_1\bnu_2
(1+\zeta^2)\frac{1+\zeta^{N-2}}{1+\zeta^N}
\big[ \tanh2\beta_2 E - \tanh2\beta_1 E \big].
\label{resjav}
\eeq
This is our `direct' result for the average instantaneous energy current density, 
valid in a finite periodic chain. 
Let ${\cal Q}$ stand for the net total  energy (i.e. time-integrated energy current), expressed in units of $4E$, that  during a time interval $[0,\tau]$ passes through the
system from thermostat 1 to thermostat 2. 
We will let $\bjm\equiv {\cal Q}/N\tau$ stand for the dimensionless
integrated current per site and per unit of time.
In the long-time limit  $\la \bjm \ra = \la \jinst\ra$
and $\la {\cal Q}\ra$ diverges with the time $\tau$ as
\beq
\la {\cal Q} \ra \simeq \la \jinst\ra N\tau, \qquad \tau\to\infty,
\label{Javdirect}
\eeq
and $\la j\ra$ given by (\ref{resjav}).
There is no such 
simple method to calculate the higher order moments $\la{\cal Q}^n\ra$
for $n\geq 2$. 
The work of this paper will lead us to expressions for the cumulants $\la{\cal Q}^n\rac$.
It will confirm equation (\ref{resjav}) as a particular case.

It is of some interest to consider the linearization in temperature around
the equilibrium state where $\beta_1=\beta_2=\beta$.
Let $\beta_a = \beta + \delta\beta_a$ for $a=1,2$ and let us set 
$\delta\beta_{12}=\delta\beta_1-\delta\beta_2=-\delta T/k_BT^2$, where $T=1/k_{\scriptscriptstyle B} \beta$
($k_{\scriptscriptstyle B}$ Boltzmann constant)  and the infinitesimal temperature difference is 
$\delta T=T_1-T_2$. 
Because of the relation (\ref{defbeta}) we then have
\beq
\delta\beta_1=\bnu_2\delta\beta_{12}, \qquad 
\delta\beta_2=-\bnu_1\delta\beta_{12}. 
\label{xdeltabetai}
\eeq
Calling the linearized current $\delta j$, we obtain from (\ref{resjav}) 
\beq
\la \delta \jinst \ra = \lambda_{\rm T}\delta T, \qquad
\lambda_{\rm T} = \bnu_1\bnu_2
\frac{(1-\zeta^2)^2(1+\zeta^{N-2})}
{2(1+\zeta^2)(1+\zeta^{N})} \beta^2 E k_{\rm B}.
\label{resjavlin}
\eeq
The heat conduction coefficient $\lambda_{\rm T}$
tends to zero in both limits $\beta\to 0$  and $\beta\to\infty$, with $E$ fixed.

\section{Extended master operator: definition and diagonalization}
\label{secinjected}

\subsection{Extended master operator $\wh{\cal M}$}
\label{secextendedME}

Each spin reversal is due to either 
$\MTh(\beta_1)$ or $\MTh(\beta_2)$,
and each spin reversal involves the injection or the release 
of a quantum of energy equal to 0 or to $\pm 4E$.
Let the integers $Q_1$ and $Q_2$ denote the total energy, measured in
units of $4E$, furnished to the system by the operators 
$\MTh(\beta_1)$
and $\MTh(\beta_2)$, respectively,
in a time interval of duration $\tau$.
For $T_1>T_2$
both $Q_1$ and $-Q_2$ will have positive expectation values.
We will write $\vecQ=(Q_1,Q_2)$.
We are interested in the joint probability distribution
$P(s,\vecQ;\tau)$, which satisfies
$
\sum_{\vecQ} P(s,\vecQ; \tau) = P(s; \tau)
\label{sumAP}
$
and
the initial condition
\beq
 P(s,\vecQ; 0)=\delta_{\vecQ,\vec0} \, P(s; 0).
\eeq
Let $s^{n}$ denote the configuration obtained from $s$ by flipping the spin at
site $n$, and let  $\Delta Q_n(s)$ denote the increment in either $Q_1$  or $Q_2$
associated with the jump from $s$ to $s^n$, that is, 
$\Delta Q_n(s)=\tfrac{1}{2}s_n(s_{n-1}+s_{n+1})$. (For the reversed spin flip at site $n$, namely the jump from $s^n$ to $s$,  the increment in either $Q_1$  or $Q_2$ is  $\Delta Q_n(s^n)=-\Delta Q_n(s)$.)
The probability $P(s,\vecQ; \tau)$ then obeys the balance equation
 \bea
\frac{ \dd P(s,\vecQ; \tau)}{\dd \tau}&=& - \left[\sum_{a=1,2} \sum_{n=1}^N w_n(s;\beta_a)\right] P(s,\vecQ; \tau)
\nonumber
\\ &+&\sum_{n=1}^N w_n(s^{n};\beta_1) P(s^{n}, Q_1+ \Delta Q_n(s), Q_2; \tau)
\nonumber
\\ &+&  \sum_{n=1}^N w_n(s^{n};\beta_2) P(s^{n}, Q_1, Q_2 + \Delta Q_n(s); \tau)
\label{BalanceP}
\eea
By analogy with the representation (\ref{defketP}) of $P(s;\tau)$, we
represent the probability $P(s, \vecQ;\tau)$ by the time dependent ket 
\beq
|P(\vecQ; \tau)\ra = \sum_s P(s,\vecQ;\tau) | s\ra.
\label{defketPbis}
\eeq
We consider the Fourier transformed ket
\beq
|\hP(\vecp; \tau)\ra =  
\sum_{\vecQ}\,\, \ee^{{\rm i} \vecp\cdot \vecQ} | P(\vecQ;\tau)\ra,
\label{defketPF}
\eeq
where $\vecp=(p_1,p_2)$ with $-\pi<p_1,p_2\leq\pi$.
Upon taking the Fourier transform of 
the balance equation 
(\ref{BalanceP}) we get the evolution equation for the ket (\ref{defketPF}),
\beq
\frac{ \dd |\hP(\vecp; \tau)\ra}{\dd \tau}= \wh{\cal M}(\vecp) |\hP(\vecp; \tau)\ra,
\label{evolTFP}
\eeq
in which
\beq
\wh{\cal M}(\vecp)= \bnu_1 \wh{{\cal M}}_{\rm th}(p_1;\beta_1)+ \bnu_2
\wh{{\cal M}}_{\rm th}(p_2;\beta_2)
\label{whcalMsum}
\eeq
where, by analogy with (\ref{defMbeta}),
\beq
\wh{{\cal M}}_{\rm th}(p_a;\beta_a)=\tfrac{1}{2}\sum_{n=1}^{N} 
\big(\sigma_n^x\ee^{- \tfrac{1}{2}{\rm i} p_a \sigma_n^z (\sigma_{n-1}^z+\sigma_{n+1}^z)}-1\big)
\big[1-\tfrac{1}{2}\gamma_a\,\sigma_n^z(\sigma_{n-1}^z+\sigma_{n+1}^z)\big].
\label{defTFMbeta}
\eeq
In this expression the operator $O_n\equiv \tfrac{1}{2}  \sigma_n^z (\sigma_{n-1}^z+\sigma_{n+1}^z)$,
whose eigenvalues are $1$, $0$, and $-1$,  has the properties 
$O_n^{2k}= O_n^2 =\tfrac{1}{2} (1+\sigma_{n-1}^z\sigma_{n+1}^z)$ 
for $k\geq 1$ and $O_n^{2k+1}= O_n$ for $k\geq 0$. Hence
$\ee^{-{\rm i} p_a O_n}= 1 - ({\rm i} \sin p_a) O_n+ (\cos p_a-1) O_n^2 $. 
As a consequence expression (\ref{whcalMsum}) may be rewritten as
\bea
\wh{\cal M}(\vecp) &=& \tfrac{1}{2}\sum_{n=1}^N 
\Big[ \tfrac{1}{2}(1+\C)\sigma_n^x
-\tfrac{1}{2}\D\,\sigma_n^x\sigma_n^z(\sigma_{n-1}^z+\sigma_{n+1}^z)
\nonumber\\
&& \phantom{XXX}-\tfrac{1}{2}(1-\C)\sigma_n^x\sigma_{n-1}^z\sigma_{n+1}^z
-1
+\tfrac{1}{2}\gamma\sigma_n^z(\sigma_{n-1}^z+\sigma_{n+1}^z) \Big],
\phantom{xxx}
\label{exMp}
\eea
in which
\bea
\C(\vecp)&=&\bnu_1 [\cos p_1 -{\rm i}\gamma_1\sin p_1]+\bnu_2[\cos p_2 -{\rm i}\gamma_2\sin p_2],
\nonumber\\
\D(\vecp)&=& \bnu_1 [\gamma_1\cos p_1 -{\rm i}\sin p_1]+\bnu_2[\gamma_2\cos p_2 -{\rm i}\sin p_2]\,.
\label{exAB}
\eea
These coefficients are real when $p_1$ and $p_2$ are pure imaginary.

\subsection{Symmetrizing the master operator}
\label{secsymm}

We apply to $\wh{\cal M}(\vecp)$ a similarity transformation
and define
\beq
\wt{\cal M}(\vecp) 
= \rho_{\rm eq}^{-\frac{1}{2}}(\betast)\wh{\cal M}(\vecp)\,
\rho_{\rm eq}^{\frac{1}{2}}(\betast),  
\label{dPsymm}
\eeq
with a $\betast(\vecp)$ left to be determined in such a way that 
$\wt{\cal M}(\vecp)$ be Hermitian. 
The only nontrivial relation needed 
to find an explicit expression for (\ref{dPsymm}) is \cite{Felderhof1971-A,Felderhof1971-B}
\bea
\tilde{\sigma}_n^x (\betast)&\equiv& 
\rho_{\rm eq}^{-\frac{1}{2}}(\betast) \sigma_n^x 
\rho_{\rm eq}^{ \frac{1}{2}}(\betast)
\nonumber\\[2mm]
&=&\sigma_n^x\Big[\cosh^2\betast E + 
\sigma_{n-1}^z\sigma_{n+1}^z\sinh^2\betast E
\nonumber\\[2mm]
&& \phantom{\big[} 
+ \sigma_n^z(\sigma_{n-1}^z+\sigma_{n+1}^z)
\sinh\betast E\cosh\betast E \Big],
\label{deftsigma}
\eea
which is easily derived.
The result is that $\wh{\cal M}(\vecp)$ of equation (\ref{exMp})
becomes an expression $\wt{\cal M}(\vecp)$ which is of the same form
as (\ref{exMp}) but with $\C$ and $\D$ of equation (\ref{exAB})
replaced with $\tC$ and $\tD$,
respectively, where
\bea
\tC(\vecp,\betast) &=& 
\C(\vecp)\cosh 2\betast E - \D(\vecp)\sinh 2\betast E,
\nonumber\\[2mm]
\tD(\vecp,\betast) &=& 
\C(\vecp)\sinh 2\betast E - \D(\vecp)\cosh 2\betast E.
\label{extAtB}
\eea
We now choose $\betast$ such that the coefficient $\tD(\betast)$
of the non-Hermitian term vanishes.
This amounts to taking
\beq
\tanh 2\betast(\vecp)E = \frac{\D(\vecp)}{\C(\vecp)}
\label{eqbetastar}
\eeq
where $\C(\vecp)$ and $\D(\vecp)$ are given by (\ref{exAB}).
We see that $\betast(\vec{0})=\beta$ and that $\betast(\vecp)$ 
is real when $p_1$ and $p_2$ are pure imaginary.
As a result the symmetrized operator $\wt{\cal M}(\vecp)$ takes the form
\beq
\wt{\cal M}(\vecp)=\tfrac{1}{2}\sum_{n=1}^N \Big[ 
\tfrac{1}{2}(1+\Cstar)\sigma_n^x
-\tfrac{1}{2}(1-\Cstar)\sigma_n^x\sigma_{n-1}^z\sigma_{n+1}^z
-1+\tfrac{1}{2}\gamma(\sigma_{n-1}^z\sigma_n^z+\sigma_n^z\sigma_{n+1}^z)
\Big]
\label{Mpsigma}
\eeq
in which $\Cstar$ is given by
\beq
\Cstar(\vecp) \equiv \tC(\vecp,\betast(\vecp)).
\label{exAstarp}
\eeq
All $\vecp$ dependence of $\wh{\cal M}(\vecp)$ is seen to enter through
the single coefficient $\Cstar(\vecp)$.

After substituting (\ref{exAB}) in (\ref{extAtB}) 
and (\ref{extAtB}) in (\ref{exAstarp})
we find that this quantity may be written as
\beq
\Cstar^2(\vecp)=1-\gamma^2 + \bTheta(\vecp)
\label{xAstar}
\eeq
where
\beq
\bTheta(\vecp)=2{\sf A}\left[\cos(p_1-p_2) -1 \right]  
+ 2{\rm i}{\sf B} \sin(p_1-p_2)
\label{defTheta}
\eeq
with ${\sf A}$ and ${\sf B}$ given by (\ref{xABintro}) in the Introduction.
These coefficients
will appear again in our final
results in section \ref{Qstat}.

\subsection{Transformation to fermion operators}
\label{secfermion}

We define fermionic quasi-particles by means of the Jordan-Wigner \cite{JordanWigner1928} transformation
\bea
c_n^\dagger &=& \tfrac{1}{2}
\left[ \,\prod_{j=1}^{n-1}\sigma_j^x \right](\sigma_n^z+{\rm i}\sigma_n^y), 
\nonumber\\[2mm]
c_n         &=& \tfrac{1}{2}
\left[ \,\prod_{j=1}^{n-1}\sigma_j^x \right](\sigma_n^z-{\rm i}\sigma_n^y),
\qquad n=1,2,\ldots,N.
\label{defcdagc}
\eea
The vacuum state of these $c$-particles is the state $|1\ra$ defined in 
(\ref{statstate}).
It is now straightforward to express $\wt{\cal M}(\vecp)$ in terms of these
fermion operators. We find from (\ref{Mpsigma})
\bea
\wt{{\cal M}}(\vecp) &=& -\tfrac{1}{4}N(1-\Cstar) 
            -\tfrac{1}{2}(1+\Cstar)\sum_{n=1}^N c^\dag_n c_n \nonumber\\
&& +\tfrac{1}{2}\gamma\sum_{n=1}^N(c^\dag_n-c_n)(c^\dag_{n+1}-c_{n+1})
\nonumber\\
&& 
-\tfrac{1}{4}(1-\Cstar)\sum_{n=1}^N(c^\dag_n-c_n)(c^\dag_{n+2}-c_{n+2})
\label{xMc}
\eea
with the understanding that
the creation and annihilation operators whose indices exceed $N$ are
defined by
\bea
c^\dag_{N+m} &=& -c^\dag_m(-1)^{{\cal N}}, \nonumber\\[2mm]
c_{N+m} &=& -c_m(-1)^{{\cal N}}, \qquad m=1,2, 
\label{defcNpm}
\eea
in which ${\cal N}=\sum_{n=1}^N c^\dag_n c_n$ is the operator for the total number of
quasi-particles. 

\subsection{Diagonalizing in terms of fermion operators}

For convenience we hence restrict ourselves to even $N$.
We define fermion operators $\etadq$ and $\etaq$ by
\bea
c^\dag_n &=& N^{-1/2}\sum_q \ee^{-{\rm i}qn} \etadq\,,  \nonumber\\[2mm]
c_n      &=& N^{-1/2}\sum_q \ee^{ {\rm i}qn} \etaq\,, \qquad n=1,...,N,
\label{defetaq}
\eea
where the wavenumber $q$ runs through the $N$ values
\beq
q=\pm\frac{\pi}{N},\pm\frac{3\pi}{N},...,\pm\frac{(N-1)\pi}{N}\,.
\label{defrangeq}
\eeq
Equation (\ref{defetaq}) is easily inverted to find the $\etadq$ and $\etaq$ 
in terms of the $c^\dag_n$ and $c_n$. 
This equation guarantees the periodicity conditions (\ref{defcNpm})
in the subspace where ${\cal N}$ is {\it even}. In that subspace
equation (\ref{defetaq}) 
may also be used in (\ref{xMc}) for $n=N+1$ and $n=N+2$.
Obviously the $c$ vacuum $|1\ra$
is also the $\eta$ vacuum.

Applying transformation (\ref{defetaq}) to (\ref{xMc}) we get
\beq
\tcalM(\vecp) = -\tfrac{1}{2}N -\tfrac{1}{2}\sum_q 
\left[C_q(\Etaone) -{\rm i}D_q(\Etatwo) \right],
\label{xMeta}
\eeq
valid in the subspace with an even number ${\cal N}$ of $c$-particles.
\footnote{In the subspace with an {\it odd\,} number of $c$-particles
  $\tcalM(\vecp)$ takes a slightly different form, as discussed in detail in
  references \cite{Felderhof1971-A,Felderhof1971-B,HilhorstETAL1972,Hilhorst1973,Hilhorst1974}
We will not need that form in this work.}, 
and where the coefficients $C_q$ and $D_q$ are given by
\bea
C_q(\vecp) &=& \tfrac{1}{2}(1+\Cstar) -\gamma\cos q +\tfrac{1}{2}(1-\Cstar)\cos 2q,
\nonumber\\[2mm]
D_q(\vecp) &=& \gamma\sin q -\tfrac{1}{2}(1-\Cstar)\sin 2q,
\label{defCD}
\eea
where the $\vecp$ dependence comes in through the $q$ independent coefficient
$\Cstar(\vecp)$ defined in (\ref{exAstarp}).
Extending the approach of Ref.\,\cite{Felderhof1971-A,Felderhof1971-B} 
to nonzero $\vecp$ we define angles $\chi_q$ (that are in general complex) by
\beq
\cos\chi_q(\vecp) = \frac{C_q}{\sqrt{C_q^2+D_q^2}}\,, \qquad
\sin\chi_q(\vecp) = \frac{D_q}{\sqrt{C_q^2+D_q^2}}\,,
\label{defchi}
\eeq
and perform in the space of the pair 
$\{\eta_q,\etadmq\}$  a Bogoliubov-Valatin
\cite{Bogoliubov1958,Valatin1958} operator rotation
\bea
\xiq(\vecp) &=& \phantom{-}
(\cos\tfrac{1}{2}\chi_q)\,\etaq - {\rm i}(\sin\tfrac{1}{2}\chi_q)\,\etadmq\,,
\nonumber\\[2mm]
\xidmq(\vecp) &=& 
- {\rm i}(\sin\tfrac{1}{2}\chi_q)\,\etaq + (\cos\tfrac{1}{2}\chi_q)\,\etadmq\,. 
\label{defxi}
\eea
It is useful to note that $\chi_{-q}=-\chi_q$. Upon using (\ref{defxi})
to transform (\ref{xMeta}) to $\xi$ operators 
we find the diagonal form
\beq
\tcalM(\vecp) = -\mu_* -\sum_q\mu_q\xidq\xiq
\label{x1Mxi}
\eeq
where
\beq
\mu_q(\vecp) = \sqrt{C_q^2+D_q^2}
\label{defmu}
\eeq
and
\beq
\mu_*(\vecp) = \frac{1}{2}\sum_{q}(1-\mu_q).
\label{defmustar}
\eeq
From (\ref{defCD}) and (\ref{defmu}) it is easily seen that $\mu_q=\mu_{-q}$.
Upon combining both equations we get for $\mu_q$ the
explicit expression
\beq
\mu_q^2 = (\gamma-\cos q)^2 + \Cstar^2(\vecp)\sin^2 q,
\label{x1muq}
\eeq
with $\Cstar(\vecp)$ given by (\ref{xAstar}).
We note that 
the generally complex quantity
$\Cstar$ does not depend on $q$ and that the $\vecp$ dependence of
this diagonalization process comes in only through $\Cstar(\vecp)$.
For $\vecp=\vec{0}$ our results for $\mu_q(\vecp)$ reduces to that of Ref.\,\cite{Felderhof1971-A,Felderhof1971-B}, namely $\mu_q(\vec{0})=1-\gamma\cos q$.

A different way, useful for later, to write the eigenvalue $\mu_q(\vecp)$ is
\beq
\mu_q^2 = (1-\gamma\cos q)^2 + \bTheta(\vecp)\sin^2 q,
\label{x2muq}
\eeq
where $\bTheta$ has been defined in (\ref{defTheta}).
We observe for later use that 
\beq
\mu_*(\vec{0})=0, \qquad \bTheta(\vec{0})=0.
\label{relvecpzero}
\eeq
It is convenient to rewrite the diagonalized form 
(\ref{x1Mxi}) of the master operator as
\beq
\tcalM(\vecp) = -\tfrac{1}{2} N - \sum_{q>0} \mu_q (\Xione),
\label{x2Mxi}
\eeq
where the symmetry property $\mu_q=\mu_{-q}$ has been employed and
where, here and hence, `$q>0$' refers to the $\tfrac{1}{2} N$ 
positive values of $q$ among those given in (\ref{defrangeq}).

\section{Joint probability distribution of the  time-integrated energy currents}
\label{secaverages}

\subsection{Joint probability distribution $P(\vecQ;\tau)$ of the time-integrated energy  currents}
\label{secdefPi}

Let $P(\vecQ;\tau)$ be the probability that at time $\tau$ the time-integrated
energies furnished by the thermostats $1$ and $2$ to the system,
counted in units of $4E$, have the
values $Q_1$ and $Q_2$ respectively. Then according to (\ref{defketPbis})
this probability distribution is
given by $P(\vecQ;\tau)=\sum_s \, \la s |P(\vecQ;\tau)\ra$  and upon inverting (\ref{defketPF}) we find
\beq
P(\vecQ;\tau) = \int_{-\pi}^{\pi}\frac{\dd p_1}{2\pi}\int_{-\pi}^{\pi}\frac{\dd p_2}{2\pi}\,\ee^{-{\rm i}\vecp\cdot\vecQ}\,\,\hP(\vecp;\tau),
\label{PInverseTF}
\eeq
where $\hP(\vecp;\tau)=\sum_s \la s |\hP(\vecp;\tau)\ra$.
The evolution equation (\ref{evolTFP}) may be formally solved as
\beq
|\hP(\vecp;\tau)\ra = \ee^{{\wh{\cal M}(\vecp)} \tau} |\hP(\vecp;0)\ra,
\eeq
where $ |\hP(\vecp;0)$ is the Fourier transform of  the initial state $P(\vecQ;0)\ra$.
Our protocol will be to take for the initial configuration
the equilibrium state at an arbitrary inverse temperature $\beta_0$.
Moreover, since
at time $\tau=0$ no energy exchange has taken place yet we choose
this probability concentrated in $\vecQ=0$, that is,
$|P(\vecQ;0)\rc =\delta_{\vecQ,\vec0} |P_{\rm eq}(\beta_0)\rc$. With the definitions
(\ref{statstate}) the  probability $\hP(\vecp;\tau)$  reads
\beq
\hP(\vecp;\tau)=\la 1|\ee^{{\wh{\cal M}(\vecp)} \tau} \rho_{\rm eq}(\beta_0)|1\ra\,.
\label{PInverseTFbis}
\eeq
This expression takes advantage of the fact that 
${\wh{\cal M}}$ is block diagonal in the subspaces of fixed $\vecp$.

\subsection{Rewriting $P(\vecQ;\tau)$}
\label{secrewrPi}

According to the complete diagonalization performed in section \ref{secinjected} 
 the matrix element in the Fourier transform (\ref{PInverseTFbis}) is an expectation value in the $\eta$ vacuum and
 can be rewritten as
\beq
P(\vecp;\tau)=
\la 1|\rho_{\rm eq}^{1/2}(\betast)\,\ee^{\tcalM(\vecp)\tau}
\rho_{\rm eq}^{-1/2}(\betast)\rho_{\rm eq}(\beta_0)|1\ra,
\label{x1hPi}
\eeq
where  we have passed to the symmetrized operator 
$\tcalM(\vecp)$ and $|1\ra$ denotes  the $\eta$-vacuum.
We decompose the $\eta$-vacuum as
\beq
|1\ra = 2^{N/2}\otimes_{q>0}|0_q0_{-q}\ra,
\label{defqvac}
\eeq
where $|0_q0_{-q}\ra$ is the state in which the
quasi-particles of wavenumbers $\pm q$ are absent,
\beq
\etaq|0_q0_{-q}\ra=0, \qquad \etamq|0_q0_{-q}\ra=0,
\label{relqvac}
\eeq
and $\la 0_q0_{-q}|0_q0_{-q}\ra=1$.

It is useful to rewrite the time evolution operator (\ref{x2Mxi}) as
\beq
{\tcalM}(\vecp) = -\tfrac{1}{2} N - \sum_{q>0} \mu_{q}\mbbX_q
\label{Mrewrite}
\eeq
with
\beq
\mbbX_q = \Xione,
\label{defX}
\eeq
where we have not indicated explicitly the $\vecp$ dependence of
the $\mbbX_q$, $\xi_q^\dagger$, and $\xi_q^\dagger$ operators. 
Furthermore we may express the Hamiltonian in terms of fermion operators,
which yields
\bea
{\cal H} &=& -2E\sum_{q>0} \mathbb{H}_q\,,
\nonumber\\[2mm]
\mathbb{H}_q &=& \mathbb{A}_q\cos q + {\rm i}\mathbb{B}_q \sin q,
\label{Hrewrite}
\eea
where
\bea
\mathbb{A}_q &=& \Etaone,  \nonumber\\[2mm]
\mathbb{B}_q &=& \Etatwo\,,  \nonumber\\[2mm]
\mathbb{D}_q &=& \Etafour\,,
\label{defAB}
\eea
where we included $\mathbb{D}_q$ for later reference.
We now use the fact that $\mbbX_q$ 
[in view of relations (\ref{defX}) and (\ref{defxi})]
and $\mathbb{H}_q$ [in view of (\ref{Hrewrite}) and (\ref{defAB})]
are quadratic in the $\eta$ operators and that therefore
\beq
[\mbbX_q,\mbbX_{q^\prime}]=
[\mbbX_q,\mathbb{H}_{q^\prime}]=
[\mathbb{H}_q,\mathbb{H}_{q^\prime}]=0,
\qquad q\neq q^\prime.
\label{relcommut}
\eeq
Upon using (\ref{defqvac}) in (\ref{x1hPi}) 
we may factorize  $\hP(\vecp;\tau)$ according to
\beq
{\hP}(\vecp;\tau)= \frac{2^N}{Z(\beta_0)}\,\ee^{-\frac{1}{2}N\tau} 
\prod_{q>0}\Zq(\vecp;\tau)
\label{factPi}
\eeq
in which
\beq
\Zq(\vecp;\tau)= \la 0_q0_{-q}|\ee^{\betast E\,\mathbb{H}_q}
\,\ee^{-\mu_q\mbbX_q\tau}
\,\ee^{(2\beta_0-\betast)E\,\mathbb{H}_q}|0_q0_{-q}\ra,
\label{defMqp}
\eeq
Since $\exp(K\mathbb{H}_q)$ (for $K=\betast E$ or $K=(2\beta_0-\betast)E$) 
and $\exp(-\mu_q\mbbX_q\tau)$ are both quadratic in the fermion operators,
they act in the two-dimensional space spanned by the vacuum $|0_q0_{-q}\ra$
defined above and the two-particle state $|1_q1_{-q}\ra$ defined by
\beq
|1_q1_{-q}\ra=\etadqdmq|0_q0_{-q}\ra.
\label{def1q}
\eeq

To make the action of $\exp(K\mathbb{H}_q)$ more explicit we expand the
exponential using the relations
\beq
\mathbb{H}_q^2 = -\mathbb{A}_q+2\mathbb{D}_q, \qquad 
\mathbb{H}_q^3 = \mathbb{H}_q, \qquad  \mathbb{H}_q^4=\mathbb{H}_q^2,
\label{relH}
\eeq
which are easily checked. 
One then obtains
\bea
\ee^{K\mathbb{H}_q} &=& 1 + \mathbb{H}_q\sinh K + \mathbb{H}_q^2(\cosh K -1)
\nonumber\\[2mm]
&=& d_0(K) + d_1(K)(\Etaoneplus) + d_2(K)(\Etatwo) \nonumber\\[2mm] 
&& + d_4(K)\Etafour
\label{expH}
\eea
in which
\bea
d_0(K) &=& \cosh K - \cos q\sinh K,      \nonumber\\[2mm]
d_1(K) &=& 1 - \cosh K + \cos q \sinh K, \nonumber\\[2mm]
d_2(K) &=& {\rm i}\sin q \sinh K,        \nonumber\\[2mm]
d_4(K) &=& 2(\cosh K - 1).
\label{defdi}
\eea
For two specific choices of $K$ we will use below the notation
\beq
b_i=d_i([2\beta_0-\betast]E), \qquad c_i=d_i(\betast E), \qquad i=0,1,2,4.
\label{defbc}
\eeq
We have to similarly expand $\exp(-\mu_q\mbbX_q\tau)$ and obtain along the
same lines
\bea
\ee^{-\mu_q\mbbX_q\tau} &=& 1 -\mbbX_q\sinh\mu_q\tau 
+ \mbbX_q^2(\cosh\mu_q\tau - 1)
\nonumber\\[2mm]
&=&  \ee^{\mu_q\tau} - (\ee^{\mu_q\tau}-1)(\Xioneplus)
+ 2(\cosh\mu_q\tau - 1)\Xifour.
\nonumber\\
&&
\label{exp1M}
\eea
With the aid of the relations (\ref{defxi}) 
we can turn this into an expansion of the form
\bea
\ee^{-\mu_q\mbbX_q\tau} &=& a_0 + a_1(\Etaoneplus) \nonumber\\[2mm]
&& + a_2(\Etatwo) + a_4\Etafour.
\label{expM}
\eea
After a fair amount of algebra 
one finds for the coefficients $a_i$ the expressions 
\bea
a_0 &=&     \cosh\mu_qt + \cos\chi_q \sinh\mu_qt, 
\nonumber\\[2mm]
a_1 &=& 1 - \cosh\mu_qt - \cos\chi_q \sinh\mu_qt, \nonumber\\[2mm]
a_2 &=& {\rm i}\sin\chi_q \sinh\mu_qt, \nonumber\\[2mm]
a_4 &=& 2(\cosh\mu_qt - 1).
\label{defai}
\eea
We substitute now expansions (\ref{expH}) 
for $K=\betast E$ or $K=(2\beta_0-\betast)E$ and (\ref{expM}) in 
(\ref{defMqp}).
Taking into account that creation (annihilation) operators acting to
the left (to the right) on the $\eta$-vacuum
give zero, we may suppress the corresponding terms in the expansions
and can write (\ref{defMqp}) as
\bea
\Zq(\vecp;\tau) &=& \la0_q0_{-q}|[c_0+c_2\etaqmq] \nonumber\\[2mm]
&& \times[a_0+a_1(\Etaoneplus)+a_2(\Etatwo)+a_4\Etafour] \nonumber\\[2mm]
&& \times[b_0+b_2\etadqdmq]|0_q0_{-q}\ra \nonumber\\[2mm]
&=& a_0b_0c_0 - a_2(b_0c_2+b_2c_0) - (a_0+2a_1+a_4)b_2c_2. 
\label{x1Mpq}
\eea
In the last line each term correspond to a sequence of creations and
annihilations as one reads the first line from the right to the left,
starting from and ending up in the vacuum.
We may now substitute the values of the $a_i$, $b_i$, and $c_i$ found above.

After some algebra applied to (\ref{x1Mpq})
we may cast the $\Zq(\vecp;\tau)$ in the form
\beq
\Zq(\vecp;\tau) = S_q(\beta_0)\left[\cosh\mu_q(\vecp)  \tau +  \frac{ T_q(\vecp;\beta_0)}{S_q(\beta_0)}\frac{\sinh\mu_q(\vecp)\tau}{\mu_q(\vecp)} \right],
\label{x2Mpq}
\eeq
where $\mu_q$ is given by (\ref{x2muq}) while
\bea
S_q (\beta_0) &=& \cosh(2\beta_0E) -  \sinh(2\beta_0E) \cos q\,, \nonumber\\[2mm]
T_q(\vecp; \beta_0) &=& S_q (\beta_0)(1-\gamma \cos q )+ U(\vecp;\beta_0) \sin^2 q   \,
\label{defSTUV}
\eea
with
\beq
U(\vecp;\beta_0)=\bnu_1 \, u(p_1; \beta_0-\beta_1) + \bnu_2 \,u(p_2; \beta_0-\beta_2)
\eeq 
and
\beq
u(p_a; \beta)=\cosh(2\beta E + {\rm i} p_a) -\cosh(2\beta E).
\eeq
Combining (\ref{factPi}) and (\ref{x2Mpq}) we get
\beq
\hP(\vecp;\tau)=\frac{2^N}{Z(\beta_0)}\,\ee^{-\frac{1}{2}N\tau}\,
\left[ \prod_{q>0}S_q(\beta_0) \right]
\prod_{q>0}\left[ \cosh\mu_q \tau + \frac{T_q(\vecp;\beta_0)}{S_q(\beta_0) \mu_q(\vecp)}\sinh\mu_q \tau\right].
\label{x2hPi}
\eeq
where the partition function defined in (\ref{defcalP}) reads
\beq
Z(\beta_0)=2^N \left[ \cosh^N \beta_0 E + \sinh^N \beta_0 E \right].
\eeq
It is easy to verify the relation
\beq
\frac{2^N}{Z(\beta_0)}\,\prod_{q>0}S_q (\beta_0)= 1.
\label{normcond}
\eeq
Using it in (\ref{x2hPi}) and substituting (\ref{x2hPi}) in 
(\ref{PInverseTF}) we finally obtain
\bea
P(\vecQ;\tau) &=& \ee^{-\frac{1}{2}N\tau}\int_{-\pi}^{\pi}\frac{\dd p_1}{2\pi}\int_{-\pi}^{\pi}\frac{\dd p_2}{2\pi}\,
\ee^{-{\rm i}\vecp\cdot\vecQ}\,    
\prod_{q>0}\left[ \cosh\mu_q \tau + \frac{T_q(\vecp;\beta_0)}{S_q(\beta_0) \mu_q(\vecp)}\sinh\mu_q \tau\right].
\nonumber\\
&&
\label{resfinPi}
\eea
This expression depends on the initial inverse temperature $\beta_0$
through the ratio $T_q(\vecp ;\beta_0)/S_q(\beta_0)$.
It is possible to show with the aid of considerable algebra that

\beq
\frac{T_q(\vec{0};\beta_0)}{\mu_q(\vec{0})} = S_q(\beta_0),
\eeq
which together with (\ref{x2hPi}), (\ref{normcond}), and (\ref{relvecpzero})
for $\mu_*(\vec{0})$ implies that  $\hP(\vec{0};\tau)=1$, equivalent to
the normalization condition $\sum_{\vecQ}P(\vecQ;\tau)=1$.

\subsection{Finite time fluctuation relation for $P(\vecQ;\tau)$}
\label{secFRone}

One can check  on the explicit expression (\ref{resfinPi}) that $P(Q_1,Q_2;\tau)$ obeys a finite time  fluctuation relation:
 by virtue of the relation $\ln \sqrt{({\sf A} +{\sf B})/({\sf A}- {\sf B})}=2(\beta_2-\beta_1)E$, the ratio of probabilities for opposite values of the couple $(Q_1,Q_2)$ is given at any  time by
\beq
\frac{P(Q_1,Q_2;\tau)}{P(-Q_1,-Q_2;\tau)}=\ee^{-4E[(\beta_1-\beta_0)Q_1+(\beta_2-\beta_0)Q_2]}.
\label{FRoneBis}
\eeq

In fact this relation relies on two key properties. First one can define the extended  transition rates associated with the extended master operator ${\cal M}_\textrm{ext}$ such that $\dd P(s,\vecQ; \tau)/ \dd \tau = \sum_{s',\vec{Q'}} \la s,\vecQ| {\cal M}_\textrm{ext}| s',\vec{Q'}\ra P(s',\vec{Q'}; \tau)$ and whose explicit expression is derived from the balance equation 
(\ref{BalanceP}). These extended  transition rates are defined between  two  triplets, each of which involves a spin configuration together with the two energies received from  thermostats since the beginning of the considered history of the system: when the system is in spin configuration $s$ and the spin at site $n$ is flipped by thermostat $a$ (with $a=1,2$) they read
$w_n^{(a)} \left(s, \Delta Q^{(1)}_n(s), \Delta Q^{(2)}_n(s)\right)= w_n(s;\beta_a)$ where the expression for 
$\Delta Q^{(a)}_n(s)=\Delta Q_n(s)$ is given  before  (\ref{BalanceP}), that for the other thermostat $b$ is $\Delta Q^{(b)}_n(s)=0$, and the transition rate between spin configuration $w_n(s;\beta_a)$ is  given  in  (\ref{defwn}). These  extended transition rates have the symmetry property obeyed by the  transition rates $w_n(s;\beta_a)$ for the two reversed transitions $s\to s_n$ and $s_n\to s$
\beq
\frac{w_n^{(a)} \left(s, \Delta Q^{(1)}_n(s), \Delta Q^{(2)}_n(s)\right)}
{w_n^{(a)} \left(s_n, -\Delta Q^{(1)}_n(s), -\Delta Q^{(2)}_n(s)\right)}
=\ee^{-\beta_a \Delta Q^{(a)}_n(s)}
\eeq
This symmetry can be considered as an extension of the
 so-called  generalized detailed balance 
\footnote{Several terminologies can be found in the literature :  ''local'' detailed balance \cite{LebowitzSpohn1999,Seifert2012},  "generalized"  detailed balance \cite{Derrida2007} or "modified" detailed balance \cite{Kurchan1998,CornuBauer2013I}.}
which involves only the  transition rates between two configurations, and where the values of 
$\Delta Q^{(1)}_n(s)$ and $\Delta Q^{(2)}_n(s)$ are determined  solely by the transition $s\to s_n$ (which is the case when a spin at a given site can be flipped by only one thermostat).
The second key property arises from the considered protocol  and the specificity of the  stationary  configuration probability in the model. Indeed  the initial spin configuration distribution is  the stationary  configuration probability at the effective inverse temperature $\beta_0$ (with a nonvanishing mean current from thermostat $1$ to thermostat $2$). Besides in the present model the latter  configuration probability is  the canonical equilibrium distribution at inverse temperature $\beta_0$.  The two key properties altogether allow one to apply usual arguments for the derivation of fluctuation relations. In the present case the precise argument is a mere transposition of that to be found for instance in \cite{CornuBauer2013I,BauerCornu2015} where  a transition between two spin configurations is  caused by only one thermostat.

As in the case where the generalized detailed balance is met by the mere transition rates between spin configurations, the property (\ref{FRoneBis}) can be interpreted in terms of some time-integrated entropy variation  as follows.
When a thermostat at inverse temperature $\beta_a$  gives an energy $4EQ_a$ to the system, its entropy variation is $\Delta S_a=-4E\beta_aQ_a$.   The exchange contribution $\Delta_{\rm exch}^{\beta_1,\beta_2}S$ to  the entropy variation of the system is  defined as  $-(\Delta S_1 + \Delta S_2)$, namely the opposite of the sum of the entropy variations of the two thermostats ; hence  $\Delta_{\rm exch}^{\beta_1,\beta_2}S=4E(\beta_1Q_1+\beta_2Q_2)$. As in Ref.\cite{CornuBauer2013I,BauerCornu2015}  we introduce the  excess exchange entropy variation of the system $\Delta_{\rm exch}^{\rm excs, \beta_0} S$ which is defined  as the difference between the exchange entropy variation under the non-equilibrium external constraint $\beta_1\neq \beta_2$ and its value under the equilibrium condition $\beta_1=\beta_2=\beta_0$ and for the same values of the energies $4EQ_1$ and $4EQ_2$  received by the system: 
$\Delta_{\rm exch}^{\rm excs, \beta_0} S=\Delta_{\rm exch}^{\beta_1,\beta_2}S-\Delta_{\rm exch}^{\beta_0,\beta_0}S$. It also reads
\beq
\Delta_{\rm exch}^{\rm excs, \beta_0} S=4E\left[ (\beta_1-\beta_0)Q_1 +(\beta_2-\beta_0)Q_2\right].
\eeq
Hence the fluctuation relation (\ref{FRoneBis}) can be rewritten as 
$
P(Q_1,Q_2;\tau)=\\ \exp[-\Delta_{\rm exch}^{\rm excs, \beta_0} S]P(-Q_1,-Q_2;\tau)
$.
As a consequence,  the probability of the  excess exchange entropy variation obeys a finite time fluctuation relation which takes  the  ``universal'' form 
\beq
P(\Delta_{\rm exch}^{\rm excs, \beta_0}S ;\tau)=\ee^{-\Delta_{\rm exch}^{\rm excs, \beta_0} S \,}P(-\Delta_{\rm exch}^{\rm excs, \beta_0}S ;\tau).
\label{FRoneS}
\eeq

\section{Statistics of the time-integrated energy current}
\label{Qstat}

\subsection{Distribution $P({\cal Q};\tau)$ of the time-integrated energy current}
\label{secP}

We now restrict our interest to the time-integrated current that
during a time interval $[0,\tau]$ 
has traversed the system. It is defined as
\beq
{\cal Q}=\tfrac{1}{2}\left(Q_1-Q_2\right),
\label{defQ}
\eeq
which may be integer or half-integer.
It measures, in units $4E$,
half the energy furnished to the system by thermostat 1 plus half the energy
extracted from it by thermostat 2. Since for long times no energy can
accumulate in the system, this quantity is, in the long time limit, equal to
the time-integrated energy current.
The particular definition (\ref{defQ}) 
is motivated by the fact that it is antisymmetric
under exchange of the two thermostats, which makes subsequent calculations
easier.

Let $P({\cal Q};\tau)$ be the probability  of ${\cal Q}$ at time $\tau$.
This marginal probability of $P(\vecQ;\tau)$ is obtained as
$P({\cal Q};\tau)=\sum_{Q_1,Q_2}\delta_{Q_1-Q_2,2{\cal Q}}\,P(\vecQ;\tau)$. 
We will from here on,
for any $\vecp$\, dependent quantity $X(\vecp)$, employ the notation 
${X}(p,-p) \equiv X^\star(p)$. 
From (\ref{resfinPi})  and the preceding definitions we then get
\beq
P({\cal Q};\tau)=
\int_{-\pi}^{\pi}\frac{\dd p}{2\pi}\,\ee^{-2{\rm i}p{\cal Q}}\,{\hP^\star}(p;\tau),
\label{defhPip}
\eeq
in which
\beq
{\hP^\star}(p;\tau)= \ee^{-\tfrac{1}{2}N\tau}
\prod_{q>0}\left[ \cosh \mu^\star_q(p) \tau
  + \frac{T^\star_q(p;\beta_0)}{S_q(\beta_0) \mu^\star_q(p)} \sinh \mu^\star_q(p) \tau \right].
\label{x1PQ}
\eeq
We observe that  $P({\cal Q};\tau)$  given by (\ref{defhPip}) and (\ref{x1PQ}) still depends on the
initial state parameter $\beta_0$. 
For the choice $\beta_0=\beta$ the system is in a stationary state for all
$\tau \geq 0$; for $\beta_0\neq\beta$ it will asymptotically tend to that state.

We take advantage of the analyticity in $p$ of the integrand 
$\hP^\star(p;\tau)$
to point out that the moment generating function
of $P({\cal Q};\tau)$, defined as $\la \ee^{\lambda {\cal Q}} \ra \equiv
\sum_{{\cal Q}} \,\ee^{\lambda {{\cal Q}}} P({{\cal Q}};\tau)$, 
exists for all real $\lambda$ and is given  by
\beq
\la \ee^{\lambda {\cal Q}} \ra =\hP^\star\left(- \frac{{\rm i} \lambda}{2};\tau\right).
\label{ExpMGF}
\eeq

\subsection{Cumulants of ${\cal Q}$ in the long-time limit}
\label{xcumulants}

In the long-time limit the cumulants per site and unit of time $\la {\cal Q}^n\rac/N\tau$ 
of the time-integrated energy  current per site and unit of time are obtained
 from the scaled cumulant generating function $g_N(\lambda)$ defined as
 \beq
Ng_N(\lambda)\equiv\lim_{\tau\to \infty} \frac{1}{\tau} \ln 
\la \ee^{\lambda {\cal Q}}\ra 
\label{defgN}
\eeq
The cumulants of interest are the values of the derivatives of $g_N(\lambda)$ with respect to $\lambda$ taken at $\lambda=0$,
\beq
\frac{\la {\cal Q}^n\rac}{N\tau} = 
\frac{\dd^n g_N(\lambda)}  {\dd \lambda^n} \Bigg|_{\lambda=0}.
\label{cumulantsDef}
\eeq

According to (\ref{ExpMGF}) and (\ref{defgN}) together with the explicit value (\ref{x1PQ}) of $\hP^\star(-{\rm i} \lambda/2;\tau)$ we get
\beq
g_N(\lambda)=
-\frac{1}{2}+\frac{1}{N}\sum_{q>0}\mu^\star_q\left(- \frac{{\rm i}\lambda}{2}\right) 
\label{expgN1}
\eeq
The expression for  $\mu_q^\star(p)=\mu_q(p,-p)$ is given by (\ref{x2muq}) where $\Theta(\vecp)$ is defined in (\ref{defTheta}). We set $\theta (\lambda)=\bTheta(-{\rm i} \lambda/2,{\rm i} \lambda/2)$ and get
\beq
g_N(\lambda) =
-\frac{1}{2}+\frac{1}{N}\sum_{q>0} \sqrt{(1-\gamma\cos q)^2 + \theta(\lambda) \sin^2 q}
\label{expgN2}
\eeq
with
\beq
\theta(\lambda)=2{\sf A}[\cosh\lambda -1] + 2{\sf B}\sinh\lambda,
\label{recallThetap}
\eeq
in which  
${\sf A}=\bnu_1\bnu_2(1-\gamma_1\gamma_2)$ and 
${\sf B}=\bnu_1\bnu_2(\gamma_2-\gamma_1)$ depend only on the kinetic and thermodynamic parameters of the model. We notice that the expression of the scaled generating function for the time-integrated current of energy has a form similar to that for various currents of interest in the case of a system of diffusing particles with pair creation and annihilation \cite{PopkovSchutz2011}. This is due to a connection  between the model considered by these authors and an Ising spin chain with Glauber dynamics.

We notice that  expression (\ref{expgN1}) for $g_N(\lambda)$ can be obtained without knowing the explicit expression of the moment generating function $\la \ee^{\lambda {\cal Q}} \ra$ at any time $\tau$. 
Indeed, the evolution of $\la \ee^{\lambda {\cal Q}}\ra $ is Markovian, as shown by 
(\ref{x1hPi}) with $p_1=-{\rm i} \lambda$ and $p_2={\rm i} \lambda$. Hence $g_N(\lambda)$ is the largest eigenvalue of $\tcalM(-{\rm i} \lambda,{\rm i} \lambda)$ and the operator expressions (\ref{Mrewrite}) and (\ref{defX}) lead to (\ref{expgN1}).

Eventually the  cumulants of the time-integrated energy current per site and  unit of time in the long-time limit  are given by 
(\ref{cumulantsDef}) and (\ref{expgN2})
\bea
\lim_{\tau\to\infty}\frac{\la {\cal Q}\ra}{N\tau} &=& 
\tfrac{1}{2}{\sf B}\Sigma_1(N,\gamma),
\label{xfirstcum}
\\[2mm]
\lim_{\tau\to\infty}\frac{\la {\cal Q}^2\rac}{N\tau} &=& 
\tfrac{1}{2}\big[ {\sf A} \Sigma_1(N,\gamma)
- {\sf B}^2 \Sigma_2(N,\gamma) \big],
\label{xsecondcum}\\[2mm]
\lim_{\tau\to\infty}\frac{\la {\cal Q}^3\rac}{N\tau} &=& 
\tfrac{1}{2}\big[ {\sf B} \Sigma_1(N,\gamma) 
- 3{\sf A}{\sf B} \Sigma_2(N,\gamma)
+ 3{\sf B}^3 \Sigma_3(N,\gamma) \big],
\label{xthirdcum}\\[2mm]
\lim_{\tau\to\infty}\frac{\la {\cal Q}^4\rac}{N\tau} &=& 
\tfrac{1}{2}\big[ {\sf A} \Sigma_1(N,\gamma) 
- (3{\sf A}^2+4{\sf B}^2) \Sigma_2(N,\gamma)
+ 18{\sf A}{\sf B}^2 \Sigma_3(N,\gamma) \nonumber\\[2mm]
&& \phantom{\tfrac{1}{2}[}
-15{\sf B}^4 \Sigma_4(N,\gamma)\big].
\label{xfourthcum}
\eea
where we have introduced
\beq
\Sigma_n(N,\gamma) = 
\frac{2}{N}\sum_{q>0}\,\,\frac{\sin^{2n} q}{(1-\gamma\cos q)^{2n-1}}\,.
\label{defSigman}
\eeq
We have indicated explicitly the dependence of $\Sigma_n(N,\gamma)$ on the size $N$ and the effective intermediate inverse temperature $\beta$ of the stationary state; we recall that $\beta$ is defined in terms of the parameter $\gamma$ through (\ref{defbeta}).The $\Sigma_n(N,\gamma)$ are monotonically increasing with $\gamma$.

Expressions for higher order cumulants may be derived by increasing
algebraic effort. 
Expression (\ref{xfirstcum}) for the time-averaged energy current has to
coincide with Eqs.\,(\ref{resjav})-(\ref{Javdirect}) 
of section \ref{secdirect}. Upon inserting the explicit expressions
for both one obtains the identity
\beq
\Sigma_1(N,\gamma)  =\frac{2}{N}\sum_{\ell=1}^{N/2} 
\frac{\sin^2 \frac{(2\ell-1)\pi}{N}}
{1-\gamma\cos\frac{(2\ell-1)\pi}{N}}
\,=\,\frac{(1+\zeta^2)(1+\zeta^{N-2})}{2(1+\zeta^N)}\,, \qquad
N=2,4,6\ldots,
\label{ident1}
\eeq
where we recall that $\zeta=\tanh\beta E$ while
$\gamma=\tanh 2\beta E$.
Eq.\,(\ref{ident1}) may be checked by explicit calculation.
It shows that $\frac{1}{2}\leq \Sigma_1(N,\gamma) \leq 1$.
We have not found similarly simple expressions for the $\Sigma_n(N,\gamma)$ 
with $n\geq 2$.
\vskip 0.3cm

The expressions for cumulants, of which the first four
are given in (\ref{xfirstcum})-(\ref{xfourthcum}), have an interesting
structure. The $n$th cumulant is an $n$th degree polynomial in the two
variables ${\sf A}$ and ${\sf B}$ with coefficients
$\Sigma_1(N,\gamma),\ldots,\Sigma_n(N,\gamma)$. The variables ${\sf A}$ and 
${\sf B}$ depend on both thermostat temperatures $T_1$ and $T_2$ but are
independent of the system size $N$. On the contrary 
the  coefficients $\Sigma_n(N,\gamma)$ vary with the system size $N$,
but depend only on the intermediate effective temperature $T$ and not on $T_1$
and $T_2$ separately.
\footnote{Recall that  $T$ depends also on the kinetic parameters $\nu_1$ and $\nu_2$\,.}
We will analyze the $\Sigma_n(N,\gamma)$ in detail in the limit of large $N$ and low effective temperature $T$ in section \ref{secInfiniteSize}.

When the two thermostats have equal temperatures, $T_1=T_2$,
one has ${\sf B}=0$. Then
only the even cumulants are nonzero,
as must be the case when one considers the energy transfer between two
thermostats at the same temperature. The even cumulants with $n\geq 4$ do not vanish:  
when $T_1=T_2$ the distribution of ${\cal Q}$ is an even but non-Gaussian function
\cite{CornuBauer2013}.

For a two-spin system
($N=2$ and $q=\pm\pi/2$) we have that $\Sigma_n(2,\gamma)=1$ for all $n$
and $\gamma$, and when expressions (\ref{xfirstcum})-(\ref{xfourthcum}) are rewritten in 
 dimensionful time $t=\tau/(\nu_1+\nu_2)$
 it appears that   the cumulants  \textit{per lattice site} $\tfrac{1}{2}\la {\cal Q}^n\rac/t$, with ${\cal Q}=\tfrac{1}{2}(Q_1-Q_2)$, are equal to the cumulants  $\la Q_1^n\rac/t$ for a pair in the model considered by Cornu and Bauer \cite{CornuBauer2013}. (In other words, in dimensionful time, when $N=2$ the scaled generating function  for cumulants \textit{per lattice site},  $(\nu_1+\nu_2) g_2(\lambda)$, is equal to the  the scaled cumulant generating function 
$\alpha(\lambda)$ for the pair of model.)
Indeed, in their model where each spin is reversed by only one
thermostat, an increment in  ${\cal Q}$  is invariant under global flip of the two spins in the initial configuration of a transition, while 
in the present model where each spin is reversed by both thermostats, an increment in  ${\cal Q}$ is
invariant under the left-right exchange of the two spins in the $N=2$ chain.

\subsection{Large deviation function of the time-integrated current  ${\cal Q}/\tau$}
\label{secsaddle}

The energy ${\cal Q}$ which goes through the system from thermostat 1 to thermostat 2 during a given time $\tau$ is determined by the whole history of the successive changes of spin configurations. 
We consider the time-integrated current  per site and per unit of time (in multiples of  $4E$), 
$\bjm$, defined by
\beq
{\cal Q}=\bjm N\tau,
\label{defq}
\eeq
According to definition (\ref{defQ}) this variable takes
the discrete values $\bjm_m=m/(2N\tau)$, where $m$ is an integer. 
As time increases the number of discrete values $\bjm_m$ in a given interval $[\bjm-\varepsilon,\bjm+\varepsilon]$ (with $\varepsilon>0$) becomes larger and larger. Then the variable $\bjm$ is said to satisfy a large deviation principle if there exists a function ${\cal I}_N(\bjm)$ such that
\footnote{For a precise discussion of this definition see Ref.\cite{CornuBauer2013I} section 5 and Appendix E.}
\beq
\lim_{\varepsilon\to 0}\lim_{\tau\to \infty} -\frac{1}{N\tau}\ln P\left(\frac{{\cal Q}}{N\tau} \in [\bjm-\varepsilon,\bjm+\varepsilon]  \right)
= {\cal I}_N(\bjm).
\label{defIN}
\eeq
The limit ${\cal I}_N(\bjm)$ is the so called large deviation function of the current $\bjm$. It vanishes for the most probable value of $\bjm$, namely when $\bjm$ is equal to $\lim_{\tau\to \infty} \la {\cal Q} \ra/N\tau$. This value coincides with the mean instantaneous current  per site in the stationary state, $\lim_{\tau\to \infty}  \la {\cal Q} \ra/N\tau=\la \jinst\ra$.

One might try to evaluate the large deviation
function ${\cal I}_N(\bjm)$ from the definition (\ref{defIN}) by considering $P({\cal Q};\tau)$ and  applying the saddle point method   to its inverse Fourier transform representation (\ref{defhPip})-(\ref{x1PQ}) rewritten as an integral on the unit circle by setting $z=\ee^{{\rm i}p}$.
However this method is mathematically tricky because of the singularities  in the complex  $z$-plane. This is exemplified by the explicit calculation of the leading behavior of  $P(\bjm_2;\tau)$ for the time-integrated current $\bjm_2$ received from  thermostat $2$  in the case of a  two spin model. We point out that the limit $\tau\to \infty$ must be taken under the  condition that  $\bjm_2 \tau$ takes only integer values (see section 6 of Ref.\cite{CornuBauer2013}).  

A far simpler method relies on the Gärtner-Ellis  theorem, which ensures that, under weak hypotheses which are fulfilled in the generic case,
the expression for ${\cal I}_N(\bjm)$ can be derived  from the sole knowledge of the  scaled cumulant generating function  $g_N(\lambda)$ defined in  (\ref{defgN}). For a Markovian process the  determination of $g_N(\lambda)$ is reduced  to  the calculation of the largest eigenvalue of the operator that governs the evolution of the generating function  $\la \ee^{\lambda {\cal Q}} \ra =\hP^\star(-{\rm i} \lambda/2;\tau)$.
 In the present case the scaled  cumulant generating function $g_N(\lambda)$,  defined in  (\ref{defgN}), exists 
and  is differentiable for all real $\lambda$, as shown by its expression (\ref{expgN2})-(\ref{recallThetap}). Thus 
$g_N(\lambda)$ satisfies the hypothesis of the simplified version of the
Gärtner-Ellis theorem (see, {\it e.g.,}
\cite{Touchette2009,DemboZeitouni1998}). 
This  version guarantees that
the large deviation function 
${\cal I}_N(\bjm)$
of the time-integrated energy current per site exists and  can be  calculated as
the Legendre-Fenchel transform of $g_N(\lambda)$, that is,
\beq
 {\cal I}_N(\bjm) =\max_{\lambda \in \mathbb{R}}\{ \lambda \bjm - g_N(\lambda)  \}.
 \label{LFTransform}
\eeq
Moreover in the present case $g_N(\lambda)$ 
is strictly convex and continuously differentiable for all real $\lambda$. As a consequence the maximum in the definition of the Legendre-Fenchel transform may be calculated with the aid of the Legendre transform,
\beq
{\cal I}_N(\bjm) = \bjm \lambda_{\bjm} - g_N(\lambda_{\bjm}),
\label{xINt}
\eeq
where $\lambda_{\bjm}$ is the solution of 
the extremum equation $\dd g_N(\lambda)/\dd\lambda= \bjm$.

In the present case this extremum
cannot be solved analytically except for the case $N=2$. Indeed,  when $N=2$ only one wave number $q=1/\pi$ is involved in the expression (\ref{expgN2}) for $g_N(\lambda)$ and the corresponding expression $g_2(\lambda)$ happens to coincide with the scaled cumulant generating function $g(\lambda)$ for another  two spin model considered by Cornu and Bauer.
Various explicit expressions of ${\cal I}_2(\bjm)$, together with some properties, can be found in section 6.1.2 of  Ref.\cite{CornuBauer2013}.
\vspace{3mm}

We point out that ${\cal I}_N(\bjm)$ obeys a  generic fluctuation relation which relies on the ratio of transition rates for two reversed jumps of configurations. It can be retrieved for the explicit solution  of the paper
 in various ways. First,  since $\ln\sqrt{({\sf A} +{\sf B})/({\sf A}- {\sf B})}=2(\beta_2-\beta_1)E$, the scaled generating function $g_N(\lambda)$ given by (\ref{expgN2})-(\ref{recallThetap}) has the symmetry property 
\beq
g_N(\lambda)=g_N\big( -\lambda-4(\beta_2-\beta_1)E \big).
\label{symmfNz}
\eeq
As a consequence ${\cal I}_N(\bjm)$ obeys the fluctuation relation
\beq
{\cal I}_N(\bjm)={\cal I}_N(-\bjm)+4(\beta_2-\beta_1)E\bjm.
\label{INfluctrel}
\eeq
This relation also appears for a system of particles moving along a line between two thermostats at different temperatures and endowed with the kinetics of a simple  exclusion process \cite{BodineauDerrida2007}.
We notice that the symmetry property (\ref{symmfNz}) determines the value of the large deviation function for $\bjm=0$. Indeed expression (\ref{LFTransform}) for the large deviation function together with (\ref{symmfNz}) implies that for zero current
the minimum is located at the point of
symmetry of $g_N(\lambda)$, namely $\lambda_0=-2(\beta_2-\beta_1)E$. As a
consequence ${\cal I}_N(0)=-g_N\big( -2(\beta_2-\beta_1)E \big)$. We notice that, since the system as a finite number of energy levels the long time fluctuation relation (\ref{INfluctrel}) for ${\cal Q}$ can be derived from the finite time fluctuation relation (\ref{FRoneBis}) for the couple of variables $Q_1$ and $Q_2$.

\subsection{Infinite size chain at finite effective temperature}
\label{InfiniteChainFiniteT}

When the system size goes to infinity  at finite effective temperature ($N\to\infty$ with $\gamma<1$), the limit of the generating function $g_N(\lambda;\gamma)$ given by (\ref{expgN2})-(\ref{recallThetap}) reads
 \bea
&&\lim_{N\to\infty} g_N(\lambda; \gamma)=-\frac{1}{2}
\\  \nonumber 
&& +  \frac{1}{2\pi} \int_{0}^{\pi} \dd q\,\sqrt{(1-\gamma\cos q)^2+ 2\left[{\sf A}(\cosh\lambda-1) + {\sf B}\sinh\lambda\right] \sin^2q}.
 \label{limgNTer}
 \eea
The  function $\lim_{N\to\infty}g_N(\lambda;\gamma)$, as well as its first derivative with respect to $\lambda$, are well defined  for all real values of $\lambda$.
Therefore, according to Gärtner-Ellis theorem,
when $N$ goes to infinity, there exists a  large deviation function ${\cal I}(\bjm ; \gamma)$ given by
$
 {\cal I}(\bjm ; \gamma) =\lim_{N\to\infty} {\cal I}_N(\bjm; \gamma)  
$.

Moreover, not only the first long time cumulant per site and unit of time $\lim_{\tau\to\infty} \la {\cal Q}\ra/N\tau$, but also all other cumulants with $n\geq 2$ remain finite in the limit of infinite size at finite effective temperature. Indeed, when $\gamma<1$,  all derivatives of $\lim_{N\to\infty} g_N(\lambda; \gamma)$ with respect to $\lambda$ have a finite value at $\lambda=0$ in this limit. 
The fact that all cumulants  per site and  unit of time remain finite in this limit can be also retrieved from the structure of the cumulants exhibited by the expressions (\ref{xfirstcum}) -(\ref{xfourthcum}) for cumulants of order $n=1,2,3,4$).  Indeed the $n$th cumulant  is a polynomial of order $n$ in the variables ${\sf A}$ and ${\sf B}$ with coefficients proportional to the $\Sigma_p(N,\gamma)$ with $p\leq n$.
The finite values of ${\sf A}$ and ${\sf B}$ are  independent of $N$  while  if $\gamma<1$
 \beq
\lim_{N\to\infty} \Sigma_n(N,\gamma) = \frac{1}{\pi}
\int_{0}^{\pi} \dd q\,\,\frac{\sin^{2n}q}{(1-\gamma\cos q)^{2n-1}}
\label{limSigma}
\eeq
is  finite for all $n\geq 1$.

\section{Various physical effects}
\label{Physicaleffects}

\subsection{Kinetic effects}
\label{seckinetic}

We call `kinetic' those effects that are related to the  
kinetic parameters $\bnu_1$ and $\bnu_2$ governing the mean frequencies of the spin flips by each thermostat. It is of interest to consider, at arbitrary fixed temperatures $T_1$ and
$T_2$\,, the condition $\nu_2/\nu_1\ll 1$. That is, the colder thermostat flips any spin more slowly than the hotter one. 
We restore in the discussion below the dimensionful physical time
variable $t=\tau/(\nu_1+\nu_2)$.
Upon expanding $(\nu_1+\nu_2)g_N$ as given by (\ref{expgN2}) and (\ref{recallThetap})
in a power series in $\nu_2/\nu_1$ we find that 
\beq
(\nu_1+\nu_2)g_N(\lambda) = \frac{\nu_2}{2}\left\{
\left[p_+ \ee^{\lambda} + p_- \ee^{-\lambda} -(p_++p_-)\right]
\Sigma_1(N,\gamma_1) 
+ {\cal O}\left(\frac{\nu_2}{\nu_1}\right)\right\},
\label{frw}
\eeq
 in which
\beq
p_+=\tfrac{1}{2}(1-\gamma_1)(1+\gamma_2),
\qquad
p_-=\tfrac{1}{2}(1+\gamma_1)(1-\gamma_2).
\label{xppm}
\eeq
The argument $\gamma_1$ of the function $\Sigma_{N,1}$ in (\ref{frw})
is the leading order
term of the expansion of $\gamma$ for small $\nu_2/\nu_1$.

The leading order term in (\ref{frw}) 
is in fact the scaled generating function 
for the cumulants of a {\it biased random walk\,} with step rates $p_+ \Sigma_1(N,\gamma_1) $ to
the right and $p_-\Sigma_1(N,\gamma_1) $ to the left, and dimensionful kinetic parameter $\nu_2$ \cite{CoxMiller1965,vanKampen1992}. 
The corresponding formulae in the case where $\nu_1\ll \nu_2$ are obtained by exchanging $\nu_1$ and $\nu_2$ and replacing $\gamma_1$ by $\gamma_2$. In other words, if  the indices $\textrm{f}$ and $\textrm{s}$ denote the fast and slow thermostats, respectively, then the scaled generating function given in (\ref{frw}) takes the generic form
\beq
(\nu_1+\nu_2)g_N(\lambda) = \frac{\nu_\textrm{s}}{2}\left\{
\left[p_+ \ee^{\lambda} + p_- \ee^{-\lambda} -(p_++p_-)\right]
\Sigma_1(N,\gamma_\textrm{f}) 
+ {\cal O}\left(\frac{\nu_\textrm{s}}{\nu_\textrm{f}}\right)\right\},
\label{frwBis}
\eeq
From the generic relation (\ref{cumulantsDef}) cumulants read to leading order in $\nu_\textrm{s}/\nu_\textrm{f}$ \bea
\nonumber
\lim_{t\to \infty}\frac{\la {\cal Q}^{2m-1}\rac}{Nt} &=&
\frac{\nu_\textrm{s} }{2}\left[(\gamma_2-\gamma_1)\Sigma_1(N,\gamma_\textrm{f}) 
+ {\cal O}\left(\frac{\nu_\textrm{s}}{\nu_\textrm{f}}\right)\right],
\\[2mm]
\label{CumRW}
\lim_{t\to \infty}\frac{\la {\cal Q}^{2m}\rac}{Nt} &=& \frac{\nu_\textrm{s}}{2} \left[(1-\gamma_1\gamma_2)\Sigma_1(N,\gamma_\textrm{f}) 
+ {\cal O}\left(\frac{\nu_\textrm{s}}{\nu_\textrm{f}}\right)\right],
\eea
for $m=1,2,\ldots$.
The latter expressions, with $t$ in place of $\tau$, can be retrieved from our expressions  (\ref{xfirstcum})-(\ref{xfourthcum}) by multiplying them by  $\tau/t=\nu_1 +\nu_2$ and expanding them to leading order in $\nu_\textrm{s}/\nu_\textrm{f}$.

\subsection{One thermostat  at zero temperature}
\label{secZeroTtwo}

Dissipation towards a thermostat at zero temperature was studied
by Farago and Pitard \cite{FaragoPitard2007,FaragoPitard2008}
for an Ising chain in which the energy is injected at a single site.
We consider here the corresponding limit for the present model.

Let thermostat 2 have $T_2=0$ while we keep $T_1>0$. Consequently $\gamma_2=1$,
which for 
${\sf A}$ and ${\sf B}$ given by (\ref{xABintro})
implies that ${\sf A} = {\sf B} = 1-\gamma_1$. Combined with
(\ref{recallThetap}) this yields
$
{\theta}(\lambda) = 2\bnu_1\bnu_2(1-\gamma_1)[\ee^{\lambda}-1].
$
When the latter expression is substituted in (\ref{expgN2}), we get that when $\gamma_2=1$
\beq
g_N(\lambda)=
-\frac{1}{2}+\frac{1}{N}\sum_{q>0} \sqrt{(1-\gamma\cos q)^2 + 2\bnu_1\bnu_2(1-\gamma_1)\left[\ee^{\lambda}-1\right] \sin^2 q}.
\label{expgN2ZeroTtwo}
\eeq
The
function $g_N(\lambda)$ is now monotonous increasing on the whole real $\lambda$
axis. It follows that the saddle point equation 
$\dd g_N(\lambda)/\dd \lambda=\bjm$ 
has no solution for $\bjm<0$, which may be restated as 
 \beq
 {\cal I}_N(\bjm)=\infty, \qquad \bjm<0.
 \eeq
This expresses the strict impossibility for the energy to flow
from the thermostat at $T_2=0$ to the one at finite temperature $T_1>0$.

The calculation of the cumulants in section \ref{xcumulants} nevertheless
remains valid and their expressions now simplify. 
The cumulants now become polynomials in $\bnu_1\bnu_2(1-\gamma_1)$.
For instance, the first two cumulants (\ref{xfirstcum}) and
(\ref{xsecondcum}) now read 
\bea
\lim_{\gamma_2\to 1}\lim_{t\to \infty}
\frac{\la Q\ra}{{N t}}&=& \frac{\nu_1+\nu_2}{2}\bnu_1\bnu_2(1-\gamma_1)\Sigma_{N,1}(\gamma),
\label{dissipationcum}
\\[2mm]
\nonumber
\lim_{\gamma_2\to 1}\lim_{t\to \infty}
\frac{\la {\cal Q}^2 \rac}{{N t}} &=& 
 \frac{\nu_1+\nu_2}{2}\bnu_1\bnu_2(1-\gamma_1)
\left[\Sigma_{N,1}(\gamma) -\bnu_1\bnu_2(1-\gamma_1)\Sigma_2(N,\gamma) \right],
\eea
where the inverse temperature $\beta$ in the special case $\gamma_2=1$ is  determined from
$\tanh2\beta E =\gamma =1-\bnu_1(1-\gamma_1)$.

\subsection{Kinetic effects when colder thermostat is at zero temperature}

When the  colder thermostat is at zero temperature, $T_2=0$, and one thermostat is faster than the other, the scaled generating function  given by  Eq.\,(\ref{frw})  becomes
\beq
(\nu_1+\nu_2)g_N(\lambda) = \frac{\nu_\textrm{s}}{2}\left\{
(1-\gamma_1)[ \ee^{\lambda} -1]\
\Sigma_1(N,\gamma_\textrm{f}) 
+ {\cal O}\left(\frac{\nu_\textrm{s}}{\nu_\textrm{f}}\right)\right\}.
\label{frw2}
\eeq
 This is the generating function for a {\it Poisson process.} 
As is well known, its cumulants are all equal, and indeed we find,
to leading order in $\nu_\textrm{s}/\nu_\textrm{f}$,
\beq
\lim_{t\to \infty}\frac{\la Q^{n}\rac}{Nt}= \frac{1}{2}\nu_\textrm{s}
\left[(1-\gamma_1)\Sigma_1(N,\gamma_\textrm{f}) 
+ {\cal O}\left(\frac{\nu_\textrm{s}}{\nu_\textrm{f}}\right)\right]
\label{nu2ll1Tzero}
\eeq
for $n=1,2,\ldots$.
By comparing (\ref{dissipationcum}) and (\ref{nu2ll1Tzero}) one sees that the limits $T_2\to 0$ and $\nu_\textrm{s}\ll \nu_\textrm{f}$ commute.

\section{Large size and low effective temperature}
\label{secInfiniteSize}

\subsection{Parameters at low effective temperature}

We now consider the regime where $N\gg1$ and $0<1-\gamma\ll1$. According to the relation $\gamma=\gamma_2-\nu_1(\gamma_2-\gamma_1)$ the condition  $0<1-\gamma\ll1$  corresponds to 
\beq
0\leq 1-\gamma_2\ll1
\eeq
 while
\beq
 0<\gamma_2-\gamma_1\ll 1\quad\textrm{and/or}\quad 0<\bnu_1 \ll 1.
\eeq
We notice that  in the case $\gamma_1=\gamma_2$  and $0\leq 1-\gamma_2\ll1$ the stationary state would correspond  to an equilibrium  state at very low temperature.

In view of later analysis  we rewrite ${\sf A}$ and ${\sf B}$, defined in (\ref{xABintro}), as
\beq
{\sf A}=(1-\gamma)\, \bnu_1\bnu_2 \,{\sf a},
\qquad
{\sf B}=(1-\gamma)\, \bnu_1\bnu_2\, {\sf b}\, ,
\label{defab}
\eeq
where ${\sf a}=(1-\gamma_1\gamma_2)/(1-\gamma)$ and ${\sf b}=(\gamma_2-\gamma_1)/(1-\gamma)$.
The model is defined for $\bnu_1\bnu_2 \neq 0$  and the non-equilibrium condition  reads $\gamma_1 <\gamma_2$. As a result the  identity  $\gamma=\gamma_2-\bnu_1(\gamma_2-\gamma_1)$   entails the hierarchy $\gamma_1\gamma_2\leq\gamma_1<\gamma<\gamma_2\leq 1$, and  $0<{\sf b}\leq {\sf a} <1$. 
\vskip0.3cm

For the sake of conciseness, from now on we denote the long time cumulants per site and unit of time
\beq
\kappa^{(n)}(N,\gamma)=\frac{1}{N}\lim_{t\to\infty}\frac{\la {\cal Q}^n\rac}{(\nu_1+\nu_2) t}.
\label{defkappatilde}
\eeq
The cumulants can be conveniently split into two contributions : a random walk process with the same first two cumulants as for the ${\cal Q}$ process and a deviation from it. The cumulants $\kappa^{(n)}$ for the random walk are denoted by $\kappa^{(n)}_\textrm{RW}$. All even (odd) cumulants take the same value, as exemplified by (\ref{CumRW}) in the case of two thermostats whose kinetic parameters are of  different orders of magnitude. The cumulant of order $n$ can be written as
\beq
\kappa^{(n)}_\textrm{RW}(N,\gamma)=(1-\gamma)\,\,\Sigma_1(N,\gamma)\,\,
 \bk_n,
\label{DefKappaRW}
\eeq
with the definition
\beq
\bk_n= \frac{1}{2} \bnu_1\bnu_2 \left[\frac{1+(-1)^n}{2}{{\sf a}} + \frac{1-(-1)^n}{2}{{\sf b}}\right],
\label{defc1}
\eeq
where ${\sf a}$ and ${\sf b}$ are defined in (\ref{defab}).
As illustrated by the expressions (\ref{xfirstcum})-(\ref{xfourthcum}) for the first four cumulants, 
the generic expression of the cumulants $\kappa^{(n)}$ are  related to those for the  corresponding  random walk as follows. The first cumulant of the ${\cal Q}$ process can be reduced to the random wall contribution 
 \beq
\kappa^{(1)}(N,\gamma)=\kappa^{(1)}_\textrm{RW}(N,\gamma),
\label{rescaleKappa1}
\eeq
while for $n\geq 2$ 
\beq
\kappa^{(n)}(N,\gamma)=\kappa^{(n)}_\textrm{RW}(N,\gamma) + \Delta \kappa^{(n)}(N,\gamma)
\label{decomp}
\eeq
where the deviation $ \Delta \kappa^{(n)}(N,\gamma)$ from the random walk process reads
\beq
\Delta \kappa^{(n)}(N,\gamma)=  \sum_{p=2}^{n}(1-\gamma)^p \,\Sigma_p(N,\gamma)  (\bnu_1\bnu_2 )^p{\sf c}_p^{(n)}({\sf a}, {\sf b}),
\label{rescaleKappan}
\eeq
In (\ref{rescaleKappan}) the factor  ${\sf c}_p^{(n)}({\sf a}, {\sf b})$ is a linear combination of terms ${\sf a}^q {\sf b}^{p-q}$, with $q=0,\ldots,p$, where the numerical coefficients  depend on the order $n$  of the cumulant; it is determined from the definition (\ref{cumulantsDef}) for  every  cumulant  per site and unit of time and the expression (\ref{expgN2})-(\ref{recallThetap}) for their  generating function.

\subsection{Finite chain at zero effective temperature}

For a finite size chain, the limit of zero effective temperature  for the scaled cumulant generating function, $\lim_{\gamma\to 1}g_N(\lambda ;\gamma)$,  is a finite sum   given  by (\ref{expgN2})-(\ref{recallThetap}) with $\gamma$ equal to one. This function  and all its derivative with respect to $\lambda$ are well defined for all real values of $\lambda$. As a consequence, when $\gamma\to 1$ all cumulants are finite and the large deviation function  exists and is given  by
${\cal I}_N(\bjm ; 1)=\lim_{\gamma\to1} {\cal I}_N(\bjm; \gamma)$.

The random walk contribution to the cumulant $\kappa^{(n)}$ is defined in (\ref{DefKappaRW}).
According to the explicit expression (\ref{ident1}) for  $\Sigma_1(N,\gamma)$, its value at $\gamma=1$ is merely
$\Sigma_1(N,1)=1$ for all $N$. Therefore in the limit $\gamma\to 1$ the random walk contribution 
$\kappa^{(n)}_\textrm{RW}(N,\gamma)$ vanishes as $1-\gamma$. More precisely
\beq
\lim_{\gamma\to 1} \frac{\kappa^{(n)}_\textrm{RW}(N,\gamma)}{1-\gamma}= \bk_n,
\label{FirstOrderBehaviorRW}
\eeq
where $\bk_n$ is defined in (\ref{defc1}). 

We now turn to  the ${\cal Q}$ process.  By virtue of (\ref{rescaleKappa1}) its first moment coincides with the first moment of the corresponding random walk, $\kappa^{(1)}(N,\gamma)=\kappa^{(1)}_\textrm{RW}(N,\gamma)$, and its leading behavior is the leading behavior of $\kappa^{(1)}_\textrm{RW}(N,\gamma)$ given by
(\ref{FirstOrderBehaviorRW}).
Besides, for all $n\geq 2$ the coefficient $\Sigma_n(N,\gamma)$ defined in (\ref{defSigman}), remains finite when $\gamma=1$ at $N$ fixed. Thus, according to the expression (\ref{rescaleKappan}), the deviation $\Delta \kappa^{(n)}(N,\gamma)$ of a cumulant from  the  corresponding random walk expression vanishes as $(1-\gamma)^2$ when $\gamma\to 1$,
\beq
\Delta \kappa^{(n)}(N,\gamma)\underset{\gamma\to 1}={\cal O}\left((1-\gamma)^2\right).
\eeq
Eventually the leading $(1-\gamma)$-term in the cumulant $\kappa^{(n)}$ is equal to the $(1-\gamma)$-term in the corresponding random walk contribution 
$\kappa^{(n)}_\textrm{RW}$. By virtue of (\ref{FirstOrderBehaviorRW}) it reads
\beq
\lim_{\gamma\to 1} \frac{\kappa^{(n)}(N,\gamma)}{1-\gamma}= \bk_n.
\label{FirstOrderBehavior}
\eeq

We point out that  the  deviation  of the first moment  $\kappa^{(1)}(N,\gamma)$ from its leading  contribution of order $1-\gamma$, denoted by $\Delta \kappa^{(1)}(N,\gamma)$,
vanishes as $(1-\gamma)^2$, as it is the case for 
 the deviation $\Delta \kappa^{(n)}(N,\gamma)$ of every higher order cumulant from the corresponding random walk cumulant. Indeed, by virtue of (\ref{rescaleKappa1}), the  difference $\Delta \kappa^{(1)}(N,\gamma)$ is the difference between the first moment of the random walk and its leading $(1-\gamma)$ term, and according to the expression (\ref{DefKappaRW}) for the random walk cumulant, it reads
 \beq
 \Delta \kappa^{(1)}= (1-\gamma) \left[\Sigma_1(N,\gamma)-1\right]\bk_1.
 \label{defDeltaKappa1}
\eeq
It can be rewritten in terms of a single finite sum 
\beq
 \Delta \kappa^{(1)}= (1-\gamma)^2 \,\bk_1\sum_{\ell =1}^{N/2}\Delta s_{1,\ell}(N,\gamma) 
\eeq
where the increment $\Delta s_{1,\ell}(N,\gamma)$ is written in  (\ref{defDeltas1l}). This finite sum indeed converges when $\gamma\to 1$, and 
 \beq
 \Delta \kappa^{(1)}\underset{\gamma\to 1}={\cal O}\left((1-\gamma)^2\right).
\eeq

\subsection{Infinite size chain at low effective temperature}

In the case of an infinite size chain  at finite effective temperature ($\gamma<1$), as discussed in subsection (\ref{InfiniteChainFiniteT}),  the large deviation function exists and  all cumulants are finite.
When $\gamma\to1$,  the scaled generating function for the cumulants still exists and it is differentiable for all $\lambda$. As a consequence, 
the large deviation exists and is given by
${\cal I}(\bjm )=\left.\lim_{N\to\infty} {\cal I}_N(\bjm; \gamma)\right\vert_{\gamma=1}$, while
the first cumulant remains finite. 

First we consider the double limit  $N\to \infty$ and $\gamma\to 1$ for the random walk process.
By virtue of the definition (\ref{DefKappaRW}) 
\beq
\lim_{N\to\infty,\gamma\to 1} \frac{\kappa^{(n)}_\textrm{RW}(N,\gamma)}{1-\gamma}
= \bk_n.
\label{kappaRWinfinitechain}
\eeq
where the notation for the limit is meant to emphasize the commutativity of the limits $N\to \infty$ and $\gamma\to 1$ for the leading $(1-\gamma)$-term in every random walk cumulant.
Indeed, according to the expression (\ref{ident1}), on the one hand $\lim_{N\to\infty} \Sigma_1(N,\gamma)=[1+(\tanh\beta E)^2]/2$ and  $\lim_{\gamma\to 1}\lim_{N\to\infty} \Sigma_1(N,\gamma)=1$, while, on the other hand, for all $N$ $\lim_{\gamma\to 1}\Sigma_1(N,\gamma)=1$  and $\lim_{N\to\infty} \lim_{\gamma\to 1}\Sigma_1(N,\gamma)=1$.

For the infinite chain (as for the finite chain)  the first cumulant $\kappa^{(1)}(\infty,\gamma)$ coincides with the first cumulant of the corresponding random walk  $\kappa^{(1)}_\textrm{RW}(\infty,\gamma)$ according to (\ref{rescaleKappa1}). Therefore the first cumulant in the double limit $N\to \infty$ and $\gamma\to 1$ also vanishes as $(1-\gamma)\bk_1$.
According to the decomposition (\ref{rescaleKappa1})-(\ref{decomp}), the deviation of $\kappa^{(n)}(N,\gamma)$ from the random walk process, $\Delta \kappa^{(n)}(N,\gamma)$, 
 is a linear combination of terms   
$(1-\gamma)^p\Sigma_p(N,\gamma)$ with $2\leq p\leq n$ given by  (\ref{decomp})-(\ref{rescaleKappan}). 
For $N$ finite  $\Delta \kappa^{(n)}(N,\gamma)$ vanishes as $(1-\gamma)^2$  when $\gamma\to1$. In the double limit $N\to\infty$ and $\gamma\to 1$ every $\Sigma_p(N,\gamma)$ with $2\leq p$ diverges according to its definition (\ref{defSigman}) and as can also be seen in the integral representation (\ref{limSigma}) for $\Sigma_n(\infty,\gamma)$.
Therefore the limits $\gamma\to 1$ and $N\to\infty$ cannot be taken independently of each other for the calculation of $\Delta \kappa^{(n)}$.
However $\Delta \kappa^{(n)}(N,\gamma)$ is expected to decay more slowly than $(1-\gamma)^2$ but still faster than $1-\gamma$ in an adequate scaling for $N$ and $1-\gamma$. Eventually, in the case of the infinite chain the cumulants vanish as $1-\gamma$, when $\gamma\to 1$ with the same coefficient as the random walk contribution
\beq
\lim_{N\to\infty,\gamma\to 1} \frac{\kappa^{(n)}(N,\gamma)}{1-\gamma}
= \bk_n.
\label{kappainfinitechain}
\eeq

 Now we turn to the correction to this leading $1-\gamma$ term. In the regime where $N\gg1$ and $0<1-\gamma\ll1$, for every cumulant $\kappa^{(n)}(N,\gamma)$ the correction to  the leading $(1-\gamma)$-term (\ref{FirstOrderBehavior})  
is the sum of two contributions arising from  (\ref{rescaleKappa1})-(\ref{decomp}):  on the one hand, the  correction (\ref{defDeltaKappa1}) to the leading $(1-\gamma)$-term in the first moment of the random walk defined in (\ref{DefKappaRW}),
and on the other hand,  the leading behavior  of  the deviation $\Delta \kappa^{(n)}$ defined in (\ref{rescaleKappan}).
In the double limit $N\gg1$ and $0<1-\gamma\ll1$,  the  sum $\left[1-\gamma\right]^{-1}\left[\Sigma_1(N,\gamma)-1\right]$, diverges as well as the 
$\Sigma_n(N,\gamma)$ for $n\geq 2$. Indeed, as detailed in Appendix \ref{Details}, and in the double limit $N\to\infty$ and $\gamma\to 1$  these sums diverge. 
These divergences can be controlled in  two scaling regimes which compare the increasing rates of $N$ and $[\gamma-1]^{-1}$, namely
\bea
\nonumber
&& \textrm{scaling regime $[\textrm{I}]$:} \quad N\sqrt{1-\gamma)} \to  +\infty
\\
\nonumber
&&\textrm{scaling regime $[\textrm{II}]$:} \quad N\sqrt{1-\gamma)} = \rho /\sqrt{2} \quad \textrm{with} \quad 
0\leq \rho<\infty.
\label{defregimescaling}
\eea
Eventually in  scaling regime $[\textrm{I}]$, the cumulants behave as
\beq
\kappa^{(n)}(N,\gamma)
\underset{\textrm{scl} \, [\textrm{I}]}{\sim}
\left[ 1-\gamma +  (1-\gamma)^{3/2}\,
\bnu_1\bnu_2\,F_n^{[\textrm{I}]}(\bnu_1\bnu_2; {\sf a} ,{\sf b}) \right]
\bk_n
\label{corr1rescaleKappan1}
\eeq
where $F_n^{[\textrm{I}]}(\bnu_1\bnu_2; {\sf a} ,{\sf b})$ does not vanish when $\bnu_1\bnu_2\to 0$, while 
in  scaling regime $[\textrm{II}]$
\beq
\kappa^{(n)}(N,\gamma)
\underset{\textrm{scl} \, [\textrm{II}]}{\sim}
\left[ 1-\gamma  +  (1-\gamma)^{3/2}\,
\bnu_1\bnu_2\,  \rho\,F_n^{[\textrm{II}]}(\rho ,\bnu_1\bnu_2; {\sf a} ,{\sf b})\right]\bk_n .
\label{corr1rescaleKappan2}
\eeq
where $F_n^{[\textrm{II}]}(\rho, \bnu_1\bnu_2; {\sf a} ,{\sf b})$ does not vanish when $\rho=0$ or $\bnu_1\bnu_2\to0$. 
We notice that the factor $\bnu_1\bnu_2$ in (\ref{corr1rescaleKappan1}) and 
(\ref{corr1rescaleKappan2})  ensures that,  when either $\nu_1\ll \nu_2$ or $\nu_1 \gg \nu_2$ even in the scaling regimes $[\textrm{I}]$ and $[\textrm{II}]$, the leading behavior of the cumulants $\kappa^{(n)}$  per  site and unit of time is still given by that of the corresponding random walk defined in (\ref{DefKappaRW}).

\subsection{Interpretation of the scaling regimes} 

Previous results can be interpreted by introducing two physical quantities:  the relaxation time to the stationary state and  the spin correlation  length.

First we recall that in the present model the dynamics for the spin configurations of the system can be seen as a Glauber dynamics with effective kinetic parameter $\nu_1+\nu_2$ and effective inverse temperature $\beta$. Hence the stationary distribution of spin configurations is the canonical equilibrium distribution at inverse temperature $\beta$, and the evolution of the spin configurations from an initial probability distribution to the stationary one is that of a relaxation to equilibrium. 
It has been shown by Glauber \cite{Glauber1963} that in the course of this relaxation the magnetization of the whole chain decays  exponentially  to its stationary value 
over the time scale $t_\textrm{rel}=[(1-\gamma)(\nu_1+\nu_2)]^{-1}$. In  other words, $t_\textrm{rel}$ is the relaxation time to the stationary state, a characteristic of which is that  the mean magnetization is constant.
 In the limit $\gamma\to 1$, $\nu_1+\nu_2$ remains finite and the relaxation time  $t_\textrm{rel}$ goes to infinity.
Therefore it is convenient to consider the long time cumulants per site and per unit of magnetization relaxation time, namely 
\beq
\lim_{t\to\infty} \frac{\la{\cal Q}^n\rac}{t/t_\textrm{rel}}=\lim_{t\to\infty} \frac{\la {\cal Q}^n\rac}{(1-\gamma) t},
\eeq
which are referred to as  ``long time rescaled cumulants''  in the following.
According to (\ref{kappainfinitechain})  all rescaled cumulants for the whole chain scale as the system size $N$ in the low-temperature regime, 
\beq
\lim_{t\to\infty}\frac{\la {\cal Q}^n\rac}{t / t_\textrm{rel}} \underset{\xi\gg 1 \, , \,N\gg 1}{\quad \sim \quad} N \,\bk_n
\label{limKdominant}
\eeq
where the finite coefficient  $k_n$ is  a random walk cumulant given by (\ref{defc1}):  for $m\geq 1$
\bea
\bk_{2m-1}&=&\frac{1}{2} \bnu_1\bnu_2 \, {\sf b}
\\
\nonumber
\bk_{2m}&=&\frac{1}{2} \bnu_1\bnu_2 \, {\sf a}
\eea
where ${\sf a}$ and ${\sf b}$ are defined in (\ref{defab}).

Second, the correlation length $\xi$ is defined from  the  correlation $\la s_ns_{n+r}\ra$ between spins at sites $n$ and $n+r$ in the infinite size chain (limit $N\to\infty$ at fixed $\beta$) when the distance  $r$ is large.
In the present model,  the stationary state for spin configurations  is the equilibrium  state at the effective inverse temperature $\beta$ defined from $\gamma$ by $\gamma=\tanh2\beta E$. 
The equilibrium correlation $\la s_ns_{n+r}\ra$ in the  Ising chain with finite size $N$ reads $\la s_ns_{n+r}\ra = (\zeta^r+\zeta^{N-r})/(1+ \zeta^N)$ with $\zeta=\tanh\beta E$. 
When $N\to\infty$ at fixed $\beta$, it takes the form $\la s_ns_{n+r}\ra = \zeta^r$ at any distance $r$. Therefore the dimensionless  correlation length $\xi(\beta)$ in the system is 
\beq
\xi (\beta)= [-\ln \tanh \beta E]^{-1}
\label{defxicorr}
\eeq
In the low effective temperature regime $[\xi (\gamma)]^{-1}\sim  2 \ee^{-2\beta E}\left[ 1+{\cal O}\left(\ee^{-2\beta E}\right)\right]$ while $1-\gamma \sim 2 \ee^{-4\beta E}\left[ 1+{\cal O}\left(e^{-4\beta E}\right)\right]$. Therefore
\beq
\sqrt{2(1-\gamma)}=\frac{1}{\xi} +{\cal O}\left(\frac{1}{\xi^2}\right).
\label{gammaxi}
\eeq
\vskip0.3cm

In the low effective temperature regime, at leading order  all rescaled cumulants for the whole chain scale as the system size $N$ (see (\ref{limKdominant})but  the behavior of the subleading term depends on the scaling regime for $N$ and $\beta$.

 In scaling regime $[\textrm{I}]$,  the size $N$ grows much faster than $\ee^{2\beta E}$ so that $(1-\gamma)^2N\gg1$. According to (\ref{gammaxi}) the latter condition implies that in  scaling regime $[\textrm{I}]$, when the temperature decreases the correlation length $\xi(\beta)$ increases but the size $N$ of the chain grows even much faster:  $(N/\xi) \gg 1$. Then the scaling behavior (\ref{corr1rescaleKappan1}) for the whole chain can be rewritten for $n\geq 1$ as
\beq
  \lim_{t\to\infty}\frac{\la {\cal Q}^n\rac}{t / t_\textrm{rel}} 
 \underset{1 \ll \xi \ll N}{\sim}  \left[ N + \frac{N}{\xi} f_n^{[\textrm{I}]}\right] \bk_n
\eeq
where $f_n^{[\textrm{I}]}=(1/\sqrt{2})( \bnu_1\bnu_2)F_n^{[\textrm{I}]}(\bnu_1\bnu_2; {\sf a} ,{\sf b})$. The correlation length $\xi$ may be viewed as the typical size of domains of parallel spins. Thus $N/\xi$ is the typical number of domains with parallel spins or equivalently the number of domain walls, $N_\textrm{dw}$, 
\beq
\frac{N}{\xi}  \underset{\textrm{scl} \, [\textrm{I}]}{\sim}N_\textrm{dw}
\quad \textrm{with}  \quad 1\ll N_\textrm{dw} \ll N
\eeq
Eventually,  any rescaled cumulant of the whole chain grows linearly with the number of sites $N$, and in scaling $[\textrm{I}]$ the correction to this leading $N$-behavior  scales  as the number of parallel spins domains $N_\textrm{dw}$.

In scaling regime $[\textrm{II}]$, the size $N$ grows as the temperature goes to zero in such a way that $(1-\gamma)N^2 = \rho^2/2$ with $\rho$ fixed and finite, namely by virtue of (\ref{gammaxi}),
\beq
\frac{N}{\xi}  \underset{\textrm{scl} \, [\textrm{II}]}{\sim} \rho < \infty.
\eeq
Then the behavior (\ref{corr1rescaleKappan2}) of the rescaled cumulants for the whole chain can be rewritten for $n\geq 1$ as
\beq
 \lim_{t\to\infty}\frac{\la {\cal Q}^n\rac}{t / t_\textrm{rel}} 
\underset{\textrm{scl} \, [\textrm{II}]}{\sim}  \left[ N + \rho^2 f_n^{[\textrm{II}]}(\rho)\right] \bk_n
\eeq
where $f_n^{[\textrm{II}]}(\rho)=(1/\sqrt{2})( \bnu_1\bnu_2)F_n^{[\textrm{II}]}/(\bnu_1\bnu_2; {\sf a} ,{\sf b})$.
In the limit where $\rho\to 0$ the function $f_n^{[\textrm{II}]}(\rho)$ goes to a non-vanishing value. The latter  regime corresponds to 
the equilibrium at  inverse temperature $\beta$ in the limit  of  very low temperature.
Eventually, in scaling $[\textrm{II}]$ the correction to the leading $N$-behavior of every rescaled cumulant of the whole chain is a finite contribution.

The size dependence of the cumulants of particle currents has  been investigated for various exclusion processes:
 the one-dimensional symmetric simple exclusion process with open boundaries
\cite{DerridaDoucotRoche2004},  on a ring with periodic boundary conditions
 \cite{AppertDerridaETAL2008}, 
 for a one-dimensional hard particle gas on a ring or with open boundary conditions
\cite{BrunetETAL2010} 
and for the one-dimensional lattice gas model  $ABC$
in the vicinity of a phase transition \cite{GerschenfeldDerrida2011}.
In Ref.\cite{MallickProlhac2009}  the  weakly asymmetric exclusion process  on a ring has been considered 
in a scaling regime where the parameter which drives the system out of equilibrium tends to zero as the inverse of the system size;
all  cumulants of a current are calculated  at both leading order and  next-to-leading order in the size of the system.

\section{Conclusion}
\label{secconclusion}

The one-dimensional Ising chain has been for a very long time
a laboratory for developing the methods of statistical physics.
In this work we have contributed to that enterprise. 
We have considered the $N$-spin cyclic chain
with each spin coupled to two thermostats at distinct
temperatures $T_1$ and $T_2$ and a dynamics that generalizes 
the Glauber \cite{Glauber1963} model.
There appears, as expected, an energy current from the hotter to the colder
thermostat.
Our fermionization method is a direct extension of the method introduced by Felderhof for the evolution of the spin probability distribution in an Ising chain with Glauber dynamics. It has allowed us to obtain the full spectrum of eigenvalues and eigenvectors
of a master equation acting in the product space of the spin configurations
and two ``counters'' that keep track of the net energy furnished by each
individual thermostats. 

In other words, we  have calculated the statistics of the total time-integrated energy current 
 ${\cal Q}$ between the thermostats
after a given time interval $\tau$.
We found an explicit expression for
the probability distribution $P({\cal Q};\tau)$ (at arbitrary finite $N$) at any time $\tau$. In the long time limit
we exhibit the  generating function for the  long time cumulants per site and unit of time $\lim_{\tau\to\infty}\la {\cal Q}^n\ra_{\rm c}$ for the transferred energy ${\cal Q}$. Their expressions  can be determined at any order $n$.
 We notice that, since the evolution of  the joint probability $P(s,{\cal Q};\tau)$ where $s$ is the spin configuration is Markovian, the  corresponding  generating function   is equal to the largest eigenvalue of the matrix that governs the evolution of the Laplace transform of $P(s,{\cal Q};\tau)$ with respect to the variable ${\cal Q}$.
Indeed, in  models solved by fermionic techniques such as those in Refs.\cite{FaragoPitard2007,FaragoPitard2008,PopkovSchutz2011}, the large deviation of the time-integrated current $X$ of interest is obtained as the largest eigenvalue of that matrix. However in these works the Laplace transform of  $P(X;\tau)$, which describes the full statistics, is not exhibited.

The explicit solution for the long time cumulants per site and unit of time has allowed us to investigate effects specific to various regimes of the thermodynamic and kinetic parameters. The main effects are the following.
When thermostat 2 is at zero temperature, the current from thermostat 1 to thermostat 2 cannot have negative fluctuations and the large deviation function is non-zero only for positive time-integrated currents : there is pure dissipation towards the zero temperature bath. When one thermostat  is very slow with respect to the other one, the generating function for the long time cumulants of ${\cal Q}$  per site and unit of time  becomes that of a biased random walk:  all odd (even) cumulants are equal to the same value. In this asymmetric random walk  the effective kinetic parameter is that of the  slower thermostat. This effect has already been exhibited in the two spin model of Ref.\cite{CornuBauer2013}. In the present model with $N\geq 2$ spins the sole coefficient due to $N$-body effects  that does contribute to the  asymmetric random walk cumulants is $\Sigma_1(N,\gamma_\textrm{f})$, where the index $\textrm{f}$ refers to the slower thermostat: the $N$-body effects  involve only  the inverse temperature of the faster thermostat. If the colder thermostat is at zero temperature, the generating function for the long time cumulants per site and unit of time becomes that of a Poisson process  with an effective  kinetic parameter equal to that of the  slower thermostat: the random biased walk is confined to positive values of ${\cal Q}$.

In this work we have dealt only with global quantities.
However, our results allow for the calculation, in principle, of any 
quantity related to the energy currents, 
and in particular energy current-current correlation functions
at different points in space and time.
This is the subject of ongoing investigation.

\appendix
\section{Behavior of coefficients $\Sigma_n(N,\gamma)$}
\label{Details}

In order to investigate the  leading behavior of  the correction 
$\Sigma_1(N,\gamma)-1$, where  $1=\lim_{N\to\infty,\,\gamma\to 1}\Sigma_1(N,\gamma)$, as well as
the divergence of  $\Sigma_n(N,\gamma)$ for $n\geq 2$ in the double limit $N\gg1$ and $0<1-\gamma\ll1$, we consider the following finite sums of interest. First we use the property  $1=\Sigma_1(N,1)$ in order to rewrite the correction $\Sigma_1(N,\gamma)-1$ as a single sum 
\beq
\Sigma_1(N,\gamma)-1= (1-\gamma)\sum_{\ell =1}^{N/2}\Delta s_{1,\ell}(N,\gamma),
\label{SumDeltaSigma1}
\eeq
where
\beq
\Delta s_{1,\ell}(N,\gamma)=-
\frac{2}{N}\frac{  \cos q_\ell\sin^2q_\ell}{(1-\cos q_\ell)} \frac{1}{[1-\gamma\cos q_\ell]}.
\label{defDeltas1l}
\eeq
and  the discrete variable $q_\ell=(2\ell -1)\pi/N$  varies between $q_1=\pi/N$ and $q_{N/2}=\pi [1- 1/N]$.
Similarly the definition (\ref{defSigman}) can be rewritten as
$
\Sigma_n(N,\gamma) = \sum_{\ell =1}^{N/2} s_{n,\ell}(N,\gamma)
$
with $n\geq 2$ and 
\beq
s_{n,\ell}(N,\gamma)=
\frac{2}{N}\,\,\frac{\sin^{2n} q_\ell}{(1-\gamma\cos q_\ell)^{2n-1}}.
\label{defsnl}
\eeq
\vskip0.3cm

 In the double limit  where $N\to\infty$ and $\gamma\to 1$ the increments 
 defined in (\ref{defDeltas1l}) and (\ref{defsnl}) have the following behavior,
\beq
\Delta s_{1,\ell}(N,\gamma) \sim
\Delta s_{1,\ell}^\star(N,\gamma)\equiv\frac{1}{N} \frac{1}{D_\ell(N,\gamma)}
\label{defDeltaSigma1coeff}
\eeq 
and 
\beq
s_{n,\ell}(N,\gamma) \sim
s_{n,\ell}^\star(N,\gamma)\equiv \frac{2}{N} \frac{q_\ell^{2n} }{[D_\ell(N,\gamma)]^{2n-1}}
\label{defSigmacoeff}
\eeq 
with the denominator
\beq
D_\ell(N,\gamma) =1-\gamma +  \frac{1}{2} \left(\frac{(2\ell -1)\pi}{N}\right)^2.
\label{defDl}
\eeq 
At this point one has to distinguish two scaling regimes  of parameters.
\vskip0.3cm

The scaling regime $[\textrm{I}]$ corresponds to  $ (1-\gamma)N^2\gg1$.  Then we rewrite the  denominator $D_\ell(N,\gamma)$ as
\beq
D_\ell(N,\gamma) =(1-\gamma )\left[1+(q_\ell^\star)^2\right]
\eeq
with $q_\ell^\star=(2\ell -1)\pi/[\sqrt{1-\gamma} N]$. Hence, from the definition (\ref{SumDeltaSigma1}), we get that when $(1-\gamma)N^2\to \infty$ 
\beq
\Sigma_1(N,\gamma)-1\underset{\textrm{scl}\, [\textrm{I}] }{\sim} 
-2 \sqrt{2(1-\gamma)}.
\label{difsigma1scl1}
\eeq
In the same limit, for $n\geq 2$ the expression  $\sqrt{2(1-\gamma)}^{(2n-3)}\sum_{\ell =1}^{N/2}s_{n,\ell}^\star (N,\gamma)$  tends to a constant denoted as $\sigma_n^{[\textrm{I}]}$ and  
 \beq
\Sigma_n(N,\gamma)\underset{\textrm{scl} \, [\textrm{I}]}{\sim} 2 \frac{1}{[2(1-\gamma)]^{n-3/2}}
 \sigma_n^{[\textrm{I}]}\,
\label{sigmanscl1}
\eeq
 with
\beq
 \sigma_n^{[\textrm{I}]}  = \frac{2^{2(n-1)}}{\pi}\int_0^{+\infty} dq  \frac{q^{2n}}{[1 + q^2]^{2n-1}}.
 \label{defcoeffsigmascl1}
\eeq
We notice that, if  $N\to\infty$ at $\gamma<1$ fixed, then $\Sigma_n(\infty,\gamma)$ is given by (\ref{limSigma}) and it   diverges as $1/(\sqrt{1-\gamma})^{2n-3}$ as $\gamma\to 1$ with the same behavior as that given in (\ref{sigmanscl1}). In other words,  the result  from the successive limits $N\to\infty$ and then $1-\gamma \ll 1$ leads to the same divergence in $1-\gamma$ as  if one considers scaling regime $[\textrm{I}]$ where $(1-\gamma)N^2\to\infty$.
\vskip0.3cm

The scaling regime $[\textrm{II}]$ corresponds to $(1-\gamma)N^2 =\tfrac{1}{2} \rho^2$ with $\rho$ fixed. Then  the denominator  $D_\ell$ defined in (\ref{defDl}) is conveniently rewritten as
\beq
D_\ell(N,\gamma)= \frac{1}{2N^2}\left[ \rho^2 +  (2\ell -1)^2\pi^2 \right]
\eeq
In the scaling regime $[\textrm{II}]$  the series $[(1-\gamma)N]^{-1}\sum_{\ell=1}^{N/2}\Delta s_{1,\ell}^\star(N,\gamma)$  tends a constant denoted as $2 \times C_\textrm{RW}(\rho)$. Then, from  the definition (\ref{SumDeltaSigma1}), we get that
\beq
\Sigma_1(N,\gamma) - 1\underset{\textrm{scl}\, [\textrm{II}] }{\sim} -   2 (1-\gamma) N\, C_\textrm{RW}(\rho). 
\eeq
 By virtue of the relation $(1-\gamma)N^2 =\tfrac{1}{2} \rho^2$, the latter behavior can be rewritten as
\beq
\Sigma_1(N,\gamma) - 1\underset{\textrm{scl}\, [\textrm{II}] }{\sim} -   \sqrt{2(1-\gamma)} \,\rho \,C_\textrm{RW}(\rho),
\label{difsigma1scl2}
\eeq
with
\beq 
C_\textrm{RW}(\rho) =4\sum_{\ell =1}^{\infty} \frac{1}{\rho^2 + (2\ell -1)^2\pi^2}.
\label{defCRrhoW}
\eeq
In the same scaling $N^{-(2n-3)}\sum_{\ell =1}^{N/2}s_{n,\ell}^\star (N,\gamma)$  tends to a constant denoted as $2\times C_n(\rho)$,
\beq
\Sigma_n(N,\gamma)\underset{\textrm{scl} \, [\textrm{II}]}{\sim} 2\,N^{(2n-3)}\, C_n(\rho).
\label{sigmanscl2N}
\eeq
By virtue of the relation $(1-\gamma)N^2 =\tfrac{1}{2} \rho^2$, the latter behavior can be rewritten as 
\beq
\Sigma_n(N,\gamma)\underset{\textrm{scl} \, [\textrm{II}]}{\sim} 
2 \left[ \frac{\rho^2}{2(1-\gamma)}  \right]^{n-3/2} C_n(\rho),
 \label{sigmanscl2}
\eeq
with 
\beq
C_n(\rho)=\frac{(2\pi)^{2n}}{2}\sum_{\ell =1}^{\infty} \frac{(2\ell -1)^{2n}}{[\rho^2 + (2\ell -1)^2\pi^2]^{2n-1}}.
\label{defCrho}
\eeq
We notice that, if the limit $\gamma\to 1$ is taken at $N$ fixed,  then the  behavior of  $\Sigma_n(N,1)$ at large $N$ is given by that of a sum where  the $\ell^\textrm{th}$ increment $s_{n,\ell}^\star(N,1)$ has the denominator 
$D_\ell(N,1)=1/(2N^2)(2\ell -1)^2\pi^2$. Then $\Sigma_n(N,1))$ behaves as   $2 N^{2n-3} C_n(0)$ where the constant $C_n(0)$ happens to be  the value of $C_n(\rho)$ (\ref{defCrho}) taken at $\rho=0$.
 In other words,  the result  from the successive limits   $\gamma\to 1$ and then $N\gg 1$ coincides with the behavior (\ref{sigmanscl2N}) of $\Sigma_n(N,\gamma)$ in scaling $[\textrm{II}]$. In other words, the divergence in $N$  of $\Sigma_n(N,\gamma)$ when the limit $\gamma\to 1$ is taken first is the same as in the scaling regime $[\textrm{II}]$  where $\rho=N\sqrt{2(1-\gamma)}$  is fixed and then sent to  zero. 

\bibliographystyle{unsrt}

\end{document}